\documentclass[12pt]{article}
\pdfoutput=1
\usepackage[colorlinks,linkcolor=Blue,citecolor=Blue,bookmarks,bookmarksnumbered]{hyperref}
\usepackage[scaled=0.85]{helvet}
\usepackage{amsmath,amssymb,accents,mathrsfs,XoohmE}
\usepackage{graphicx,color}
\usepackage{booktabs}
\usepackage[footnotesize,bf,textfont={it},margin=1 cm]{caption}
\usepackage{setspace} 
\usepackage{multirow}
\usepackage{placeins}
\usepackage{amsmath}

\def\brL{{\bm {\rm L}}}
\def\brR{{\bm {\rm R}}}

\definecolor{Green}  {rgb}{0.10,0.70,0.10} 
\definecolor{Orange} {rgb}{1.00,0.50,0.15} 
\definecolor{Red}    {rgb}{0.90,0.00,0.12} 
\definecolor{Purple} {rgb}{0.50,0.25,0.55} 
\definecolor{Turque} {rgb}{0.00,0.65,0.85} 
\definecolor{Blue}   {rgb}{0.00,0.00,1.00} 
\definecolor{Magenta}{rgb}{1.00,0.00,1.00} 
\definecolor{Gold}   {rgb}{1.00,0.75,0.25} 
\definecolor{Seaweed}{rgb}{0.01,0.24,0.09} 
\definecolor{Brown}  {rgb}{0.43,0.26,0.32} 
\definecolor{grey1}  {rgb}{0.20,0.20,0.20} 
\definecolor{grey2}  {rgb}{0.40,0.40,0.40} 
\definecolor{grey3}  {rgb}{0.60,0.60,0.60} 
\definecolor{grey4}  {rgb}{0.80,0.80,0.80} 
\definecolor{grey5}  {rgb}{0.90,0.90,0.90} 
\def\C#1#2{{\ifcase#1\or
             \color{Green}\or \color{Orange}\or \color{Red}\or
              \color{Purple}\or \color{Turque}\or \color{Blue}\or
               \color{Magenta}\or \color{Gold}\or \color{Seaweed}\or
                \color{Brown}\or\color{grey1}\or\color{grey2}\or
                 \color{grey3}\else\color{grey4}\fi#2}}

\definecolor{Slate} {rgb}{0.00,0.45,0.55}

\definecolor{AdinkraGreen}{rgb}{0.10196079, 0.61176473, 0.21960784 }
\definecolor{AdinkraViolet}{rgb}{0.42352942, 0.15294118, 0.4509804 }
\definecolor{AdinkraOrange}{rgb}{0.89803922, 0.57647061, 0.27450982}
\definecolor{AdinkraRed}{rgb}{0.78431374, 0, 0.12156863}

\def\rD{\mbox{\textcolor{AdinkraRed}{${\rm D}_4$}}}

\def\brL{{\bm {\rm L}}}
\def\brR{{\bm {\rm R}}}
\def\brV{{\bm {\rm V}}}
\def\brtV{{\bm{\widetilde{\rm V}}}}

\def\nmSG{{$\not$mSG}}

\def\rD{{\rm D}}
\def\rI{{\rm I}}
\def\rJ{{\rm J}}
\def\rK{{\rm K}}
\def\rL{{\rm L}}

\def\fracm#1#2{\hbox{\large{${\frac{{#1}}{{#2}}}$}}}

\def\be{\begin{equation}}
\def\ee{\end{equation}}
\newcommand{\bea}{\begin{eqnarray}}
\newcommand{\eea}{\end{eqnarray}}
\newcommand{\ena}{\end{eqnarray}}

\def\pp{{\mathchoice
          {
              \kern 1pt%
              \raise 1pt
              \vbox{\hrule width5pt height0.4pt depth0pt
                    \kern -2pt
                    \hbox{\kern 2.3pt
                          \vrule width0.4pt height6pt depth0pt
                          }
                    \kern -2pt
                    \hrule width5pt height0.4pt depth0pt}%
                    \kern 1pt
           }
            {
              \kern 1pt%
              \raise 1pt
              \vbox{\hrule width4.3pt height0.4pt depth0pt
                    \kern -1.8pt
                    \hbox{\kern 1.95pt
                          \vrule width0.4pt height5.4pt depth0pt
                          }
                    \kern -1.8pt
                    \hrule width4.3pt height0.4pt depth0pt}%
                    \kern 1pt
            }
            {
              \kern 0.5pt%
              \raise 1pt
              \vbox{\hrule width4.0pt height0.3pt depth0pt
                    \kern -1.9pt  
                    \hbox{\kern 1.85pt
                          \vrule width0.3pt height5.7pt depth0pt
                          }
                    \kern -1.9pt
                    \hrule width4.0pt height0.3pt depth0pt}%
                    \kern 0.5pt
            }
            {
              \kern 0.5pt%
              \raise 1pt
              \vbox{\hrule width3.6pt height0.3pt depth0pt
                    \kern -1.5pt
                    \hbox{\kern 1.65pt
                          \vrule width0.3pt height4.5pt depth0pt
                          }
                    \kern -1.5pt
                    \hrule width3.6pt height0.3pt depth0pt}%
                    \kern 0.5pt
            }
        }}

\def\mm{{\mathchoice
                       {
                             \kern 1pt
               \raise 1pt    \vbox{\hrule width5pt height0.4pt depth0pt
                                  \kern 2pt
                                  \hrule width5pt height0.4pt depth0pt}
                             \kern 1pt}
                       {
                            \kern 1pt
               \raise 1pt \vbox{\hrule width4.3pt height0.4pt depth0pt
                                  \kern 1.8pt
                                  \hrule width4.3pt height0.4pt depth0pt}
                             \kern 1pt}
                       {
                            \kern 0.5pt
               \raise 1pt
                            \vbox{\hrule width4.0pt height0.3pt depth0pt
                                  \kern 1.9pt
                                  \hrule width4.0pt height0.3pt depth0pt}
                            \kern 1pt}
                       {
                           \kern 0.5pt
             \raise 1pt  \vbox{\hrule width3.6pt height0.3pt depth0pt
                                  \kern 1.5pt
                                  \hrule width3.6pt height0.3pt depth0pt}
                           \kern 0.5pt}
                       }}

\def\ad{{\kern0.5pt
                   \alpha \kern-5.05pt \raise5.8pt\hbox{$\textstyle.$}\kern
0.5pt}}

\def\bd{{\kern0.5pt
                   \beta \kern-5.05pt \raise5.8pt\hbox{$\textstyle.$}\kern
0.5pt}}

\def\qd{{\kern0.5pt
                   q \kern-5.05pt \raise5.8pt\hbox{$\textstyle.$}\kern
0.5pt}}
\def\Dot#1{{\kern0.5pt
     {#1} \kern-5.05pt \raise5.8pt\hbox{$\textstyle.$}\kern
0.5pt}}

\catcode`@=11
\def\un#1{\relax\ifmmode\@@underline#1\else
        $\@@underline{\hbox{#1}}$\relax\fi}
\catcode`@=12

\def\a{\alpha}
\def\b{\beta}
\def\c{\chi}
\def\d{\delta}
\def\e{\epsilon}

\def\g{\gamma}

\def\i{\iota}
\def\j{\psi}
\def\k{\kappa}
\def\l{\lambda}
\def\m{\mu}
\def\n{\nu}
\def\o{\omega}

\def\r{\rho}
\def\s{\sigma}
\def\t{\tau}

\def\z{\zeta}

\def\L{\Lambda}
\def\O{\Omega}
\def\P{\Pi}

\def\S{\Sigma}

\def\ca{{\cal A}}

\def\cf{{\cal F}}

\def\dslash{\not{\hbox{\kern-2pt $\partial$}}}
\def\Dslash{\not{\hbox{\kern-4pt $D$}}}
\def\pslash{\not{\hbox{\kern-2.3pt $p$}}}
 \newtoks\slashfraction
 \slashfraction={.13}
 \def\slash#1{\setbox0\hbox{$ #1 $}
 \setbox0\hbox to \the\slashfraction\wd0{\hss \box0}/\box0 }
 
\def\kcr{{\hbox{\ro \char'170}}}                
\def\ktl{{\hbox{\ro \char'170}}}        
\def\ktr{{\hbox{\ro \char'170}}}        
\def\kbl{{\hbox{\ro \char'170}}}        
\def\kbr{{\hbox{\ro \char'170}}}        

\def\plpl{\raise-2pt\hbox{$\raise3pt\hbox{$_+$}\hskip-6.67pt\raise0.0pt
\hbox{$^+$}\hskip 0.01pt$}}
\def\mimi{\raise-2pt\hbox{$\raise3pt\hbox{$_-$}\hskip-6.67pt\raise0.0pt
\hbox{$^-$}\hskip 0.01pt$}} 

\def\bo{{\raise.15ex\hbox{\large$\Box$}}}               
\def\pa{\partial}                                       
\def\TH{{\raise.2ex\hbox{$\displaystyle \bigodot$}\mskip-4.7mu \llap H \;}}
\def\face{{\raise.2ex\hbox{$\displaystyle \bigodot$}\mskip-2.2mu \llap {$\ddot
        \smile$}}}                                      

\def\dt#1{\on{\hbox{\bf .}}{#1}}                
\def\Dot#1{\dt{#1}}

\def\leftrightarrowfill{$\mathsurround=0pt \mathord\leftarrow \mkern-6mu
        \cleaders\hbox{$\mkern-2mu \mathord- \mkern-2mu$}\hfill
        \mkern-6mu \mathord\rightarrow$}
\def\dvec#1{\vbox{\ialign{##\crcr
        \leftrightarrowfill\crcr\noalign{\kern-1pt\nointerlineskip}
        $\hfil\displaystyle{#1}\hfil$\crcr}}}           
\def\dt#1{{\buildrel {\hbox{\LARGE .}} \over {#1}}}     

\def\fracm#1#2{\hbox{\large{${\frac{{#1}}{{#2}}}$}}}
\def\sfrac#1#2{{\vphantom1\smash{\lower.5ex\hbox{\small$#1$}}\over
        \vphantom1\smash{\raise.4ex\hbox{\small$#2$}}}} 
\def\bfrac#1#2{{\vphantom1\smash{\lower.5ex\hbox{$#1$}}\over
        \vphantom1\smash{\raise.3ex\hbox{$#2$}}}}       
\def\afrac#1#2{{\vphantom1\smash{\lower.5ex\hbox{$#1$}}\over#2}}    

\def\pa{\partial}      
\let\bm\relax
\newcommand{\bm}[1]{{\boldsymbol{#1}}}

\def\ad{{\dot{\alpha}}}
\def\bd{{\dot{\beta}}}

 \font\rOpe=cmsy10                        
 \def\ktl{{\hbox{\rOpe\char'170}}}        
 \def\kbl{{\hbox{\rOpe\char'170}}}        
 \def\kcr{{\reflectbox{\rOpe\char'170}}}        
 \def\ktr{{\reflectbox{\rOpe\char'170}}}        
 \def\kbr{{\reflectbox{\rOpe\char'170}}}        
 \def\Border{\vbox{\hsize0pt
        \setlength{\unitlength}{1mm}
        \newcount\xco
        \newcount\yco
        \xco=-21
        \yco=12
        \begin{picture}(0,0)(-7.5,0)
        \put(\xco,\yco){$\ktl$}
        \advance\yco by-1
        {\loop
        \put(\xco,\yco){$\kcr$}
        \advance\yco by-2
        \ifnum\yco>-240
        \repeat
        \put(\xco,\yco){$\kbl$}}
        \xco=170
        \yco=12
        \put(\xco,\yco){$\ktr$}
        \advance\yco by-1
        {\loop
        \put(\xco,\yco){$\kcr$}
        \advance\yco by-2
        \ifnum\yco>-240
        \repeat
        \put(\xco,\yco){$\kbr$}}
        \put(-19.5,13){\scalebox{.5835}{%
         **Brown University**Department of Physics**Brown University**Department of Physics**Brown University**Department of Physics**Brown University**Department of Physics*}}
        \put(-19.5,-241.5){\scalebox{.5835}{%
         **Brown University**Department of Physics**Brown University**Department of Physics**Brown University**Department of Physics**Brown University**Department of Physics*}}
        \end{picture}
        \par\vskip-8mm}}
\definecolor{UMred}{rgb}{.9,.05,.2}
\definecolor{HUblue}{rgb}{.0,.3,.7}

\definecolor{Red}    {rgb}{0.90,0.00,0.12} 
\definecolor{Blue}   {rgb}{0.00,0.00,1.00} 
\definecolor{Green}  {rgb}{0.10,0.70,0.10} 
\definecolor{Turque} {rgb}{0.00,0.65,0.85} 
\definecolor{Orange} {rgb}{1.00,0.50,0.15} 
\definecolor{Magenta}{rgb}{1.00,0.00,1.00} 
\definecolor{Gold}   {rgb}{1.00,0.75,0.25} 
\definecolor{Seaweed}{rgb}{0.01,0.24,0.09} 
\definecolor{Purple} {rgb}{0.50,0.25,0.55} 
\definecolor{Brown}  {rgb}{0.43,0.26,0.32} 
\definecolor{grey1}  {rgb}{0.20,0.20,0.20} 
\definecolor{grey2}  {rgb}{0.40,0.40,0.40} 
\definecolor{grey3}  {rgb}{0.60,0.60,0.60} 
\definecolor{grey4}  {rgb}{0.80,0.80,0.80} 
\definecolor{grey5}  {rgb}{0.90,0.90,0.90} 
\def\C#1#2{{\ifcase#1\or
             \color{Red}\or \color{Green}\or \color{Blue}\or\
              \color{Turque}\or \color{Orange}\or \color{Magenta}\or 
               \color{Gold}\or \color{Seaweed}\or \color{Purple}\or
                \color{Brown}\or\color{grey1}\or\color{grey2}\or
                 \color{grey3}\else\color{grey4}\fi#2}}

\definecolor{Slate} {rgb}{0.00,0.45,0.55}

\newdimen\parshift\parshift=\parindent
\catcode`@=11
 \long\def\@footnotetext#1{\insert\footins{\reset@font\footnotesize
           \interlinepenalty\interfootnotelinepenalty\splittopskip%
            \footnotesep\splitmaxdepth\dp\strutbox\floatingpenalty\@MM%
             \hsize\columnwidth\addtolength{\hsize}{-2\parindent}
              \@parboxrestore\protected@edef\@currentlabel%
              {\csname p@footnote\endcsname\@thefnmark}%
                \color@begingroup%
                 \@makefntext{\rule\z@\footnotesep\ignorespaces#1%
                  \@finalstrut\strutbox}%
                \color@endgroup}}
 \long\def\@makefntext#1{\hglue\parshift%
           \vbox{\noindent\baselineskip=11pt plus.5pt minus.5pt\hb@xt@0em{\hss\@makefnmark\kern1pt}#1}}
\catcode`@=12

\newskip\humongous \humongous=0pt plus 1000pt minus 1000pt
\def\caja{\mathsurround=0pt}
\def\eqalign#1{\,\vcenter{\openup2\jot \caja
        \ialign{\strut \hfil$\displaystyle{##}$&$
        \displaystyle{{}##}$\hfil\crcr#1\crcr}}\,}
\newif\ifdtup

\makeatletter
\def\section{\@startsection{section}{1}{\z@}
        {3ex plus-1ex minus-.2ex}{1pt plus1pt}{\large\sf\bfseries\boldmath}}
\def\subsection{\@startsection{subsection}{2}{\z@}
         {1.5ex plus-1ex minus-.2ex}{0.01pt plus1pt}{\sf\slshape}}
\def\subsubsection{\@startsection{subsubsection}{3}{\z@}
          {1.5ex plus-1ex minus-.2ex}{0.01pt plus0.2pt}{\sf\boldmath}}
\def\paragraph{\@startsection{paragraph}{4}{\z@}
           {.75ex \@plus.5ex \@minus.2ex}{-2mm}{\sf\bfseries\boldmath}}
\makeatother

\def\adinkrawidth{0.48\textwidth}

 \allowdisplaybreaks
 \seceq

\begin{document}

\thispagestyle{empty}
%
\noindent{\small
\today\hfill{
 {Brown-HET-1781 ~}
}
\vspace*{8mm}
\begin{center}
{\large \bf
4D, $\mathcal{N}=1$ Matter Gravitino Genomics}   \\   [12mm]
{\large 
 S.-N. Hazel Mak\footnote{\href{mailto:sze\_ning\_mak@brown.edu}{sze\_ning\_mak@brown.edu}}
and Kory Stiffler\footnote{\href{mailto:kory\_stiffler@brown.edu}{kory\_stiffler@brown.edu}}}
\\*[12mm]
\emph{
\centering
Department of Physics, Brown University,
\\[1pt]
Box 1843, 182 Hope Street,
Providence, RI 02912, USA 
}
 \\*[10mm]
{ ABSTRACT}\\[2mm]
\parbox{142mm}{\parindent=2pc\indent\baselineskip=14pt plus1pt

{\small 
Adinkras are graphs that encode a supersymmetric representation's transformation laws that have been reduced to one dimension, that of time. A goal of the supersymmetry ``genomics'' project is to classify all 4D, $\mathcal{N}=1$ off-shell supermultiplets in terms of their adinkras.  In~previous works, the genomics project uncovered two fundamental isomer adinkras, the cis- and trans-adinkras, into which all multiplets investigated to date can be decomposed. The number of cis- and trans-adinkras describing a given multiplet define the isomer-equivalence class to which the multiplet belongs. A further refining classification is that of a supersymmetric multiplet's holoraumy: the commutator of the supercharges acting on the representation. The one-dimensionally reduced, matrix representation of a multiplet's holoraumy defines the multiplet's holoraumy-equivalence class. Together, a multiplet's isomer-equivalence and holoraumy-equivalence classes are two of the main characteristics used to distinguish the adinkras associated with different supersymmetry multiplets in higher dimensions. This paper focuses on two matter gravitino formulations, each with 20 bosonic and 20 fermionic off-shell degrees of freedom, analyzes them in terms of their isomer-~and holoraumy-equivalence classes, and compares with non-minimal supergravity which is also a $20~\times~20$~multiplet. This analysis fills a missing piece in the supersymmetry genomics project, as now the isomer-equivalence and holoraumy-equivalence for representations up to spin two in component fields have been analyzed for 4D, $\mathcal{N}=1$ supersymmetry. To handle the calculations of this research effort, we have used the a \emph{Mathematica} software package called \href{https://hepthools.github.io/Adinkra/}{\emph{Adinkra.m}}. This package is open-source and available for download at a \href{https://hepthools.github.io/Adinkra/}{\emph{GitHub} Repository}. Data files associated with this paper are also published open-source at a \href{https://hepthools.github.io/Data/}{Data Repository} also on \emph{GitHub}. 
}
 }
 \end{center}
\vfill
\vfill
\clearpage

\section{Introduction}
Generally, there are more representations of supersymmetry (SUSY) in lower dimensions than there are in higher dimensions. Given a particular higher dimensional SUSY representation, one can always reduce the representation to lower dimension simply by considering the fields of the multiplet to depend only on the subset of coordinates required. For instance, in efforts known as ``supersymmetric genomics''~\cite{G-1,Gates:2011aa,Chappell:2012qf}, a 4D SUSY multiplet is reduced to 1D by considering the fields in the multiplet to depend only on time~$\tau$. This is known as reducing to the 0-brane and the transformation laws for the 4D, $\mathcal{N}=1$ chiral multiplet, for instance, when reduced to the 0-brane can be entirely encoded in a graph known as an adinkra~\cite{Faux:2004wb,Doran:2005zt,Doran:2006it,Doran:2008kg,Doran:2008rp,Douglas:2010am,Doran:2011gb,Zhang:2011np}, as shown in Figure~\ref{f:CMAdinkra}. The precise meaning of the lines in the 
adinkra diagram are   described in Section~\ref{s:AdinkraReview}. Adinkras have been and continue to be used to discover new representations of supersymmetry: the 4D, $\mathcal{N}=2$ off-shell relaxed extended tensor mutliplet~\cite{Doran:2007bx} and a finite representation of the hypermultiplet~\cite{Faux:2016ygh}, for instance.
\begin{figure}[!htbp]
	\centering
	\includegraphics[width = 0.48 \textwidth]{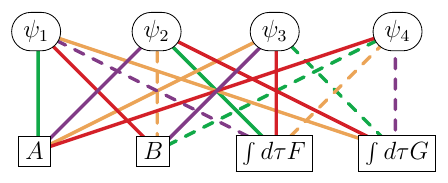}
	\caption{An adinkra for the 0-brane reduced transformation laws of the 4D, $\mathcal{N}=1$ chiral multiplet.}
	\label{f:CMAdinkra}
\end{figure}

In previous supersymmetric genomics works~\cite{G-1,Gates:2011aa,Chappell:2012qf}, the authors found adinkra representations for the 4D, $\mathcal{N}=1$ chiral multiplet (CM), vector multiplet (VM), tensor multiplet (TM), complex linear superfield multiplet (CLS), old-minimal supergravity (mSG), non-minimal supergravity (\nmSG), and conformal supergravity (cSG). They found these representations       to decompose into a number $n_c$ of fundamental cis-adinkras and a number $n_t$ of trans-adinkras, as shown in Figure~\ref{f:ctAdinkra} and tabulated in Table~\ref{t:genomicssummary} where $\chi_0 = n_c - n_t$.
\begin{figure}[!htbp]
\centering
	\includegraphics[width = \adinkrawidth]{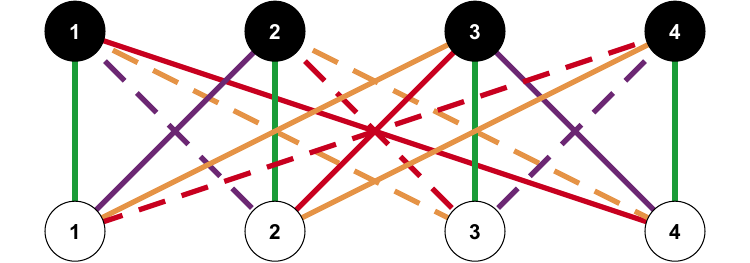} \quad \includegraphics[width = \adinkrawidth]{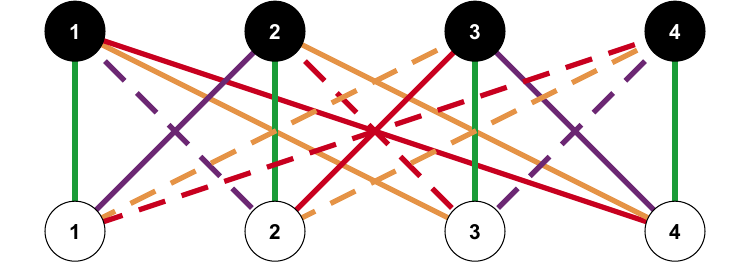}
	\caption{The cis-adinkra (\textbf{left}) and trans-adinkra (\textbf{right}). The graphs are identical aside from the orange transformation laws, which are dashed in one graph and   solid in the other. This is reflected in the ${\rm D}_3$ transformation laws which differ by a minus sign.} 
	\label{f:ctAdinkra}
\end{figure}

\vspace{-6pt}

\begin{table}[!htbp]
\centering
\caption{The number of cis-adinkras $n_c$ and trans-adinkras $n_t$ used to decompose various 4D, $\mathcal{N}=1$ multiplets and their associated value of $\chi_0 = n_c - n_t$.}
\begin{tabular}{cccccccc}
\toprule
	\textbf{Multiplet} & \textbf{CM} & \textbf{VM} & \textbf{TM} & \textbf{CLS} & \textbf{mSG} & \textbf{\nmSG} & \textbf{cSG} \\
	\midrule
	$n_c$ & 1 & 0 & 0 & 1 & 1 & 1 & 0 \\ 
	$n_t$ & 0 & 1 & 1 & 2 & 2 & 4 & 2 \\
	$\chi_0$ & $+1$ & $-1$ & $-1$ & $-1$ & $-1$ & $-3$ & $-2$\\
\bottomrule
\end{tabular}
\label{t:genomicssummary}
\end{table}

Notice in Table~\ref{t:genomicssummary} that the values of $n_c$ and $n_t$ only serve to partially differentiate the various multiplets at the adinkra level. Specifically, the CM is seen to be distinct from the VM and TM, but the VM and TM are indistinguishable based solely on their cis/trans adinkra content. Similarly, there is no difference between CLS and mSG at the adinkra-level. Another tool is needed to completely sort this~out.

Dimensional enhancement, or SUSY holography, is the effort to build higher dimensional SUSY representations from lower dimensional representations~\cite{enhanc1,enhanc2,adnkM3}. To aid in sorting out which of the multitude of lower dimensional systems are candidates for dimensional enhancement, a tool known as holoraumy is being developed~\cite{KIAS1,KIAS2,Calkins:2014exa,H4,H3,H2,G&G1,G&G2,Gates:2017tly,Gates:2017eui,Gates:2018pxb}. On the 0-brane with $N$ SUSY transformations $\rD_\rI$, holoraumy is defined as the commutator of two supersymmetric transformations $\rD_\rI$ acting on the bosons $\Phi$ or fermions $\Psi$ of a particular representation 
\begin{align}\label{e:holoraumy}
	[ \rD_\rI , \rD_\rJ ] \Phi =&  -2i \mathscr{B}_{\rm IJ} \dot{\Phi} \cr
	[ \rD_\rI , \rD_\rJ ] \Psi =& -2i \mathscr{F}_{\rm IJ} \dot{\Psi}.
\end{align}
where a dot above a field indicates a time derivative, $\dot{\Phi} = d\Phi/d\tau$ for example. Contrast holoraumy with the SUSY algebra, which is the anti-commutator of two SUSY transformations $\rD_\rI$. A closed 1D SUSY algebra takes the following form on all fields
\begin{align}\label{e:closure}
	\{ \rD_\rI , \rD_\rJ \} = 2 i \delta_{\rm IJ} \partial_\tau.
\end{align}
Holoraumy is the tool being developed to split the degeneracy of the cis- and trans-information, as shown in Table~\ref{t:genomicssummary}. For instance, holoraumy was first shown to separate the VM and TM at the adinkra level in~\cite{KIAS1,KIAS2}. To do so, a dot product-like ``gadget'' was introduced in~\cite{KIAS1,KIAS2} and studied further in~\cite{H2,G&G1,G&G2,Gates:2017tly,Gates:2017eui}.

In this paper, we further the supersymmetry genomics efforts by decomposing the two representations of 4D, $\mathcal{N}=1$ matter gravitino, one as described in~\cite{dWvH,Fradkin:1979as,MGM} and the other as described in \cite{OS1,OS2}, in terms of the cis- and trans-adinkra content as well as their holoraumy. These multiplets have the same degrees of freedom (20 bosons and 20 fermions) as \nmSG~\cite{SFSG}, thus we compare their cis- and trans-adinkra and holoraumy to this multiplet as well. To the knowledge of the authors, this paper is the first time that the transformation laws for the two matter gravitino multiplets and \nmSG~ have each been written in terms of a one-parameter family of transformation laws that encodes a field redefinition of the auxiliary fermions that preserves the diagonal character of the Lagrangian. {The~parameter is discussed in~\cite{MGM}, but the transformation laws described there are in terms of a specific value of the parameter.} Whereas the cis- and trans-adinkra content are shown to be independent of this parameter, the holoraumy is not. 
{Adinkras such as those in Figure~\ref{f:ctAdinkra} can be transformed amongst each other via signed permutations of the colors and/or nodes.  For the adinkras in Figure~\ref{f:ctAdinkra}, the group to perform these transformation is $BC_4$: the group of signed permutations of four elements. Recently, $BC_4$ transformations between adinkras of different holoraumy-equivalence classes were worked out in~\cite{Gates:2017eui} for adinkras such as those in Figure~\ref{f:ctAdinkra}.} In a sense, this paper is a sister paper of the recent work~\cite{Gates:2017eui}: this paper reviews the status of SUSY genomics, focusing on 4D dynamics, whereas~\cite{Gates:2017eui} reviews the status of SUSY holography, focusing on 1D dynamics. It is the view of at least one of the authors of this paper (KS) that building a complete picture of SUSY holography~\cite{enhanc1,enhanc2,adnkM3} would require efforts into SUSY genomics~\cite{G-1,Gates:2011aa,Chappell:2012qf}, enumeration techniques~\cite{Zhang:2018lmo,Friend:2018ree}, and classification schemes~\cite{KIAS1,KIAS2,Calkins:2014exa,H4,H3,H2,G&G1,G&G2,Gates:2017tly,Gates:2017eui,Gates:2018pxb}. {
The main results of the paper are as follows:
\begin{enumerate}
	\item This paper presents the Majorana representation of the $O(1)$ transformation laws of the multiplet described in~\cite{dWvH,Fradkin:1979as,MGM}. The existence of these transformation laws is discussed in~\cite{dWvH,Fradkin:1979as} as a submultiplet of the overarching $SO(2)$ transformation laws. In~\cite{MGM}, the $O(1)$ submultiplet's tranformation laws are presented in a Weyl representation. It is important to have a Majorana representation of component transformation laws for a multiplet to be decomposed as adinkras as in the previous genomics works~\cite{G-1,Gates:2011aa,Chappell:2012qf}.
	\item 
	
	{
	The transformation laws for the two matter-gravitino multiplets and \nmSG~ are expressed in terms of a field redefinition parameter that preserves the diagonal character of the Lagrangian. The existence of this parameter is pointed out in~\cite{MGM}. As  we plan to further SUSY genomics and holography research to higher spin multiplets, where this diagonal parameter continues to be present, it is important to understand this parameter's significance at the adinkra level.
	}
	\item This paper demonstrates the utility of the new \emph{Mathematica} package \emph{Adinkra.m} (\url{https://hepthools.github.io/Adinkra/}). 
This package is available open-source and will be indispensable in future adinkra research.
	\item The main purpose of the paper is the adinkranization of the two matter-gravitino multiplets, the~calculation of their fermionic holoraumy matrices along with that of \nmSG, and the comparisons between these three multiplets via the gadget. In calculations of holoraumy and the gadget of these three multiplets, we see the presence of the diagonal Lagrangian parameter. The~gadget results presented in this paper will provide a template for researching the significance of this parameter in future, higher spin investigations as pertaining to SUSY genomics and holography.
The $20 \times 20$ multiplets investigated in this paper are the base of a tower of higher spin multiplets~\cite{Gates:1996xs,GK1,GK2,GK3}, thus they lay the foundation for future investigations of these higher spin~multiplets.
\end{enumerate}
}

This paper is organized as follows. In Section~\ref{s:AdinkraReview}, we review adinkras. In Section~\ref{s:GenRev}, we review supersymmetry genomics. Sections~\ref{s:dWvH}--\ref{s:nmSG} review the two matter gravitino multiplets and the \nmSG~multiplet, expressed in terms of the one parameter diagonal Lagrangian family of transformations. Section~\ref{s:adinkranization} presents the adinkra and 1D holoraumy content of each of the $20 \times 20$ multiplets and makes comparisons via the gadget. For all gamma matrix conventions, we follow precisely the previous three supersymmetry genomics works~\cite{G-1,Gates:2011aa,Chappell:2012qf}.

\newpage
\section{Adinkra Review}\label{s:AdinkraReview}
Adinkras are graphs that encode supersymmetry transformation laws in one dimension (that of time $\tau$) with complete fidelity. Take for example four dynamical bosonic fields $\Phi_i$ and four dynamical fermionic fields $\Psi_{\hat{j}}$ that depend only on time. Two distinct possible sets of supersymmetry transformation laws for $\Phi_i$ and $\Psi_{\hat{j}}$ can be succinctly encoded as
\begin{subequations}
\label{e:DPhi}
\begin{align}
	\label{e:DPhi1}
	\text{${\rm D}_1$} \Phi_1 = & i \Psi_1,~~~\text{${\rm D}_2$} \Phi_1 = i \Psi_2 ,~~~\text{${\rm D}_3$} \Phi_1 = \chi_0 i \Psi_3,~~~\text{${\rm D}_4$} \Phi_1 = -i \Psi_4  \\
	\label{e:DPhi2}
	\text{${\rm D}_1$} \Phi_2 = & i \Psi_2,~~~\text{${\rm D}_2$} \Phi_2 = -i \Psi_1  ,~~~\text{${\rm D}_3$} \Phi_2 = \chi_0 i \Psi_4 ,~~~\text{${\rm D}_4$} \Phi_2 = i \Psi_3  \\
	\label{e:DPhi3}
	\text{${\rm D}_1$} \Phi_3 = & i \Psi_3,~~~\text{${\rm D}_2$} \Phi_3 = -i \Psi_4  ,~~~\text{${\rm D}_3$} \Phi_3 = -\chi_0 i \Psi_1 ,~~~\text{${\rm D}_4$} \Phi_3 = -i \Psi_2  \\
	\label{e:DPhi4}
	\text{${\rm D}_1$} \Phi_4 = & i \Psi_4,~~~\text{${\rm D}_2$} \Phi_4 = i \Psi_3  ,~~~\text{${\rm D}_3$} \Phi_4 = -\chi_0 i \Psi_2,~~~\text{${\rm D}_4$} \Phi_4 = i \Psi_1  
\end{align} 
\end{subequations}
and
\begin{subequations}
\label{e:DPsi}
\begin{align}
	\label{e:DPsi1}
	\text{${\rm D}_1$} \Psi_1 = &  \dot{\Phi}_1,~~~\text{${\rm D}_2$} \Psi_1 = -\dot{\Phi}_2  ,~~~\text{${\rm D}_3$} \Psi_1 = -\chi_0 \dot{\Phi}_3,~~~\text{${\rm D}_4$} \Psi_1 =  \dot{\Phi}_4  \\
	\label{e:DPsi2}
	\text{${\rm D}_1$} \Psi_2 = & \dot{\Phi}_2,~~~\text{${\rm D}_2$} \Psi_2 = \dot{\Phi}_1  ,~~~\text{${\rm D}_3$} \Psi_2 = -\chi_0\dot{\Phi}_4,~~~\text{${\rm D}_4$} \Psi_2 = -\dot{\Phi}_3  \\
	\label{e:DPsi3}
	\text{${\rm D}_1$} \Psi_3 = & \dot{\Phi}_3,~~~\text{${\rm D}_2$} \Psi_3 = \dot{\Phi}_4  ,~~~\text{${\rm D}_3$} \Psi_3 = \chi_0 \dot{\Phi}_1 ,~~~\text{${\rm D}_4$} \Psi_3 = \dot{\Phi}_2  \\
	\label{e:DPsi4}
	\text{${\rm D}_1$} \Psi_4 = & \dot{\Phi}_4,~~~\text{${\rm D}_2$} \Psi_4 = -\dot{\Phi}_3  ,~~~\text{${\rm D}_3$} \Psi_4 = \chi_0 \dot{\Phi}_2 ,~~~\text{${\rm D}_4$} \Psi_4 =  -\dot{\Phi}_1 
\end{align} 
\end{subequations}
where a dot above a field indicates a time derivative, $\dot{\Phi}_i = d\Phi_i/d\tau$ for example. One set of transformation laws is encoded by the choice $\chi_0=+1$ and another by $\chi_0= -1$. For Equations~(\ref{e:DPhi}) and~(\ref{e:DPsi}), there is no possible set of field redefinitions for which the $\chi_0=+1$ transformation laws reduce to the transformation laws for which $\chi_0=-1$. Owing to an analogy to isomers in chemistry, in~\cite{G-1}, the $\chi_0=+1$ transformation laws were dubbed the cis-multiplet and $\chi_0=-1$ the trans-multiplet. Both the cis- and trans-transformation laws satisfy the closure relationship, Equation~(\ref{e:closure}).

The adinkras in Figure~\ref{f:ctAdinkra} can be seen to encode the transformation laws in Equations~(\ref{e:DPhi}) and~(\ref{e:DPsi}) as~follows. 
 \begin{enumerate}
 	\item The white nodes encode the  bosons $\Phi_i$ and the black nodes encode the fermions multiplied by the imaginary number $ i \Psi_{\hat{j}}$. 
 	\item A line connecting two nodes indicates a SUSY transformation law between the corresponding~fields. 
 	\item Each of the $N=4$ colors encodes a different SUSY transformation as color coded in Equations~(\ref{e:DPhi}) and~(\ref{e:DPsi}). 
 	\item A solid (dashed) line indicates a plus (minus) sign in SUSY transformations. 
 	\item In transforming from a higher node to a lower node (higher mass dimension field to one-half lower mass dimension field), a time derivative appears on the field of the lower node. 
\end{enumerate}

The adinkras in Figure~\ref{f:ctAdinkra} are     known as \emph{valise} adinkras: adinkras with a single row of bosons and a single row of fermions. The distinction between the two set of transformation laws encoded in Equations~(\ref{e:DPhi}) and~(\ref{e:DPsi}) can be seen easily in the adinkras in Figure~\ref{f:ctAdinkra}: the two adinkras are identical aside from the orange lines, which are dashed in one adinkra and solid in the other. This is reflected in the ${\rm D}_3$ transformation laws in Equations~(\ref{e:DPhi}) and~(\ref{e:DPsi}), which differ by a minus sign.

Both  $\chi_0 = \pm 1$ supersymmetric transformation laws are symmetries of the Lagrangian
\begin{align}\label{e:L0}
	\mathcal{L} = \frac{1}{2} \delta^{ij}\dot{\Phi}_i \dot{\Phi}_j - i \frac{1}{2} \delta^{\hat{i}\hat{j}}\Psi_{\hat{i}} \dot{\Psi}_{\hat{j}}.
\end{align}
The transformation laws in Equations~(\ref{e:DPhi}) and~(\ref{e:DPsi}) can succinctly be written as
 \begin{align}\label{e:D}
	{\rm D}_{\rm I} \Phi = i \brL_{\rm I} \Psi,~~~{\rm D}_{\rm I} \Psi = \brR_{\rm I} \dot{\Phi},
\end{align}
where the adinkra matrices $\brL_{\rm I}$ are given by 
 \begin{align}\label{e:Lmatrices}
\mbox{
{
$	\brL_1 = \left(
\begin{array}{cccc}
 1 & 0 & 0 & 0 \\
 0 & 1 & 0 & 0 \\
 0 & 0 & 1 & 0 \\
 0 & 0 & 0 & 1 \\
\end{array}
\right)
$}}
&,~~~
\mbox{
{
$\brL_2 =  \left(
\begin{array}{cccc}
 0 & 1 & 0 & 0 \\
 -1 & 0 & 0 & 0 \\
 0 & 0 & 0 & -1 \\
 0 & 0 & 1 & 0 \\
\end{array}
\right)\
$}}
,\cr
\mbox{
{
$
\brL_3 = \chi_0 \left(
\begin{array}{cccc}
 0 & 0 & 1 & 0 \\
 0 & 0 & 0 & 1 \\
 -1 & 0 & 0 & 0 \\
 0 & -1 & 0 & 0 \\
\end{array}
\right)
$}}
&,~~~
\mbox{
{
$
\brL_4 = \left(
\begin{array}{cccc}
 0 & 0 & 0 & -1 \\
 0 & 0 & 1 & 0 \\
 0 & -1 & 0 & 0 \\
 1 & 0 & 0 & 0 \\
\end{array}
\right)
$}}
\end{align}
and the $\brR_\rI$ given by
\begin{align}\label{e:Linverse}
	\brR_\rI = \brL_{\rm I}^{-1}.
\end{align} 

In the specific case of the matrices~in Equation \eqref{e:Lmatrices}, we also have the orthogonality relationship
\begin{align}\label{e:adinkraic}
	\brR_\rI = \brL_{\rm I}^{-1}=\brL_{\rm I}^{T}~~~
\end{align} 
where the $T$ denotes transpose. 
Supersymmetric multiplets whose adinkra matrices satisfy the orthogonality relationship (Equation~(\ref{e:adinkraic})) are said to be \emph{adinkraic} representations, that is, they can be expressed as adinkras pictures as in Figures~\ref{f:CMAdinkra} and~\ref{f:ctAdinkra}.  
Generally, larger multiplets such as those investigated in this paper are non-adinkraic when nodes are chosen to be identified with single fields as reviewed in Sections~\ref{s:CLS} and~\ref{s:mSG}.
Non-adinkraic multiplets have been investigated previously in~\cite{Gates:2012zr,Doran:2013vea}.

The closure relation, Equation~(\ref{e:closure}), for an adinkriac system is reflected in the adinkra matrices $\brL_\rI$ and $\brR_\rI$ satisfying the $\mathcal{GR}(d,N)$ algebra also known as the garden algebra~\cite{Gates:1995ch,Gates:1995pw}
\begin{align}\label{e:GRdN}
	\brL_\rI \brR_\rJ + \brL_\rJ \brR_\rI =  2 \delta_{\rm IJ} {\bm {\rm I}}_4,~~~\brR_\rI \brL_\rJ + \brR_\rJ \brL_\rI =  2 \delta_{\rm IJ} {\bm {\rm I}}_4.
\end{align} 
The  $\mathcal{GR}(d,N)$ algebra is the algebra of general, real matrices encoding the supersymmetry transformation laws between $d$ bosons, $d$ fermions, and $N$ supersymmetries. Specifically, all the $\chi_0=\pm 1$ adinkras in Figure~\ref{f:ctAdinkra}   satisfy the $\mathcal{GR}(4,4)$ algebra. 

 For an arbitrary $d$, $N=4$ adinkra, $\chi_0$ can be defined off the following chromocharacter equation~\cite{G-1}
\begin{align}\label{e:chromo}
	Tr(\brL_{\rm I} \brR_\rJ \brL_\rK \brR_\rL) = 4[(n_c + n_t) (\delta_{\rI \rJ}\delta_{\rK \rL} - \delta_{\rI \rL}\delta_{\rJ \rK} + \delta_{\rI \rJ}\delta_{\rK \rL}) + \chi_0~\epsilon_{\rI \rJ \rK \rL}] ~~~
\end{align}
Calculating $\chi_0$ through Equation~(\ref{e:chromo}) allows one to immediately determine $n_c$ and $n_t$ which satisfy~\cite{G-1,Gates:2011aa,Chappell:2012qf}
\begin{align}
	n_c = \frac{d}{8} + \frac{\chi_0}{2} \\
	n_t = \frac{d}{8} - \frac{\chi_0}{2}
\end{align}
For $\mathcal{GR}(4,4)$ valise adinkras, the only two possible values are $\chi_0 = \pm 1$~\cite{G-1} and either $n_c = 1,~ n_t =0$ or $n_c = 0,~ n_t  = 1$.

The matrix representation of 1D holoraumy, Equation~(\ref{e:holoraumy}), is~\cite{G&G1,G&G2}
\begin{align}\label{e:Vdef}
\brV_{\rI\rJ} =& - \frac{i}{2} \textbf{L}_{[I} \textbf{R}_{J]},~~~
\brtV_{\rI\rJ} = - \frac{i}{2} \textbf{R}_{[I} \textbf{L}_{J]}.
\end{align}
where  $\brV_{\rI\rJ}$ is the matrix representation of the bosonic holoraumy tensor $\mathscr{B}_{\rm IJ}$ and $\brtV_{\rm IJ}$ is the matrix representation of the fermionic holoraumy tensor $\mathscr{F}_{\rm IJ}$. Adinkras that share the same value of $\chi_0$ and have the same number of degrees of freedom $d$ are said to be in the same $\chi_0$-equivalence class~\cite{Gates:2017eui}. There are two possible $\chi_0$-equivalence classes for $\mathcal{GR}(4,4)$ valise adinkras: the cis-equivalence class defined by $\chi_0 = +1$ and the trans-equivalence class defined by $\chi_0 = -1$.  In~\cite{Gates:2017tly}, all possible 36,864 $\mathcal{GR}(4,4)$ adinkras were investigated and tabulated and in~\cite{Gates:2017eui} all were categorized in terms of $\chi_0$-equivalence classes and holoraumy-equivalence classes.

\section{Supersymmetry Genomics Review}\label{s:GenRev}
In this section, we review the previous SUSY genomics works~\cite{G-1,Gates:2011aa,Chappell:2012qf}. In doing so, we demonstrate how the the cis-adinkra and trans-adinkra shown in Figure~\ref{f:ctAdinkra} encode the 0-brane reduced transformation laws for various 4D, $\mathcal{N}=1$ off-shell multiplets and comment on their values of $\chi_0$, holoraumy, and gadgets in the cases where this is known.

\subsection{The 4D, \texorpdfstring{$\mathcal{N}=1$}{N=1} Off-Shell Chiral Multiplet (CM)}
The dynamical field content of the CM is a scalar $A$, pseudoscalar $B$, and Majorana fermion $\psi_a$. The auxiliary field content of the CM is a scalar $F$ and pseudoscalar $G$. The 4D component Lagrangian for the CM is given by
\be\label{eq:CMLagrangian}\eqalign{
   {\mathcal L}_{\text{CM}} = & -\frac{1}{2}(\partial_{\mu}A)(\partial^{\mu}A) -\frac{1}{2}(\partial_{\mu} B)(\partial^{\mu}B)+i\frac{1}{2}(\gamma^{\mu})^{ab}\psi_{a}\partial_{\mu}\psi_{b} +\frac{1}{2} F^{2}+\frac{1}{2} G^{2} 
}\ee
The transformation laws that are a symmetry of the Lagrangian~in Equation (\ref{eq:CMLagrangian}) are
\be
\eqalign{
{\rm D}_a A  ~&=~ \psi_a ,~~~
{\rm D}_a B  ~=~ i\, ( \gamma^5 )_a{}^b \psi_b  ~~~~~~~~\,
 \cr
{\rm D}_a \psi_b  ~&=~ i\, ( \gamma^{\mu} )_{ab}  \left( \,\partial_{
\mu} A  \, \right) - ( \gamma^5 \gamma^{\mu})_{ab}  \left( \, 
\partial_{\mu} B  \, \right)  \cr
& {~~~~~} - i C_{ab}  F   ~+~ 
( \gamma^5 )_{ab}  \, G  
~~~~~~~~~~~~~~\,~~~~~~~~~~~~~~~~\,\cr
{\rm D}_a F  ~&=~ ( \gamma^{\mu})_a{}^b \,  \pa_{\mu} \psi_b, ~~~
{\rm D}_a G  ~=~ i\, ( \gamma^5 \gamma^{\mu} )_a{}^b \, \pa_{\mu} \psi_b   
~~\,~~~~~~~~~~~~~
}  \label{BV1}
\ee

Reducing to the 0-brane with nodal field definitions
\begin{subequations}
\label{e:CMnodes}
\begin{align}
	\label{e:CMfermions}
	 &  i\Psi_1 = \psi_1,~~~i\Psi_2 = \psi_2,~~~i\Psi_3 = \psi_3,~~~i\Psi_4 = \psi_4~~~ \\
	\label{e:CMbosons}
	&\Phi_1 = A,~~~\Phi_2 = B,~~~\Phi_3 = \int d\tau~F,~~~\Phi_4 = \int d\tau ~G~~~ 
\end{align}
\end{subequations} 
reduces the Lagrangian~in Equation (\ref{eq:CMLagrangian}) to Equation~(\ref{e:L0}) and transformation laws to Equation~(\ref{e:D}) with the $\brL_\rI$ matrices given by
\begin{align}\label{e:CM}
\mbox{
{
$
	\brL^{(\text{CM})}_1 = \left(
\begin{array}{cccc}
 1 & 0 & 0 & 0 \\
 0 & 0 & 0 & -1 \\
 0 & 1 & 0 & 0 \\
 0 & 0 & -1 & 0 \\
\end{array}
\right)
$}}
&,~~~
\mbox{
{
$
\brL_2^{(\text{CM})} =\left(
\begin{array}{cccc}
 0 & 1 & 0 & 0 \\
 0 & 0 & 1 & 0 \\
 -1 & 0 & 0 & 0 \\
 0 & 0 & 0 & -1 \\
\end{array}
\right)
$}}
,\cr
\mbox{
{
$
\brL_3^{(\text{CM})} =\left(
\begin{array}{cccc}
 0 & 0 & 1 & 0 \\
 0 & -1 & 0 & 0 \\
 0 & 0 & 0 & -1 \\
 1 & 0 & 0 & 0 \\
\end{array}
\right)
$}}
&,~~~
\mbox{
{
$
\brL_4^{(\text{CM})} =\left(
\begin{array}{cccc}
 0 & 0 & 0 & 1 \\
 1 & 0 & 0 & 0 \\
 0 & 0 & 1 & 0 \\
 0 & 1 & 0 & 0 \\
\end{array}
\right)
$}}.
\end{align} 
and $\brR_\rI$ given by Equation~(\ref{e:adinkraic}) and satisfy the $\mathcal{GR}(4,4)$ algebra Equation~(\ref{e:GRdN}). By the rules explained in Section~\ref{s:AdinkraReview}, it can be seen that the 0-brane transformation laws for the CM, Equation~(\ref{e:D}),  
with 
{$\brL_\rI$ and $\brR_\rI$ matrices as in Equations~(\ref{e:CM}) and~(\ref{e:adinkraic})}
and nodal field definitions~(Equation \eqref{e:CMnodes}), are entirely described by the adinkra in Figure~\ref{f:CMAdinkra2}, which is the same image as Figure~\ref{f:CMAdinkra} in the Introduction. The CM is in the cis-equivalence class as can be seen in the following two ways
\begin{enumerate}
	\item Calculating the trace in Equation~(\ref{e:chromo}), which produces the result $\chi_0 = +1$.
	\item  Performing certain field redefinitions, as in~\cite{G-1}, that transform Figure~\ref{f:CMAdinkra2} into the cis-adinkra in Figure~\ref{f:ctAdinkra}.
\end{enumerate}

The field redefinitions mentioned in the second of these are dubbed \emph{flips} and \emph{flops} in the recent work~\cite{Gates:2017eui}.

\begin{figure}[!htbp]
	\centering
	\includegraphics[width = 0.48 \textwidth]{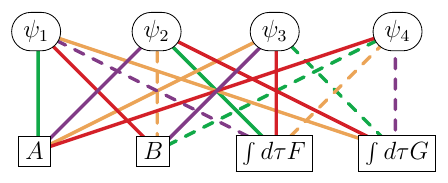}
	\caption{This is identical to Figure~\ref{f:CMAdinkra}: An adinkra for the 0-brane reduced transformation laws of the 4D, $\mathcal{N}=1$ chiral multiplet.} 
	\label{f:CMAdinkra2}
\end{figure}

\subsection{The 4D, \texorpdfstring{$\mathcal{N}=1$}{N=1} Off-Shell  Tensor Multiplet (TM)}
The dynamical field content of the TM is a scalar $\varphi$, anti-symmetric rank-two tensor $B_{\mu\nu}$, and a Majorana fermion $\chi_a$. There are no auxiliary fields in the TM. The 4D component Lagrangian for the TM is given by
\be\label{eq:TMLagrangian}\eqalign{
 \mathcal{L}_{\text{TM}} = &  - \frac{1}{3}H_{\mu\nu\alpha}H^{\mu\nu\alpha}- \frac{1}2 \partial_\mu \varphi \partial^\mu \varphi   + \frac{1}{2} i (\gamma^\mu)^{bc} \chi_b \partial_\mu \chi_c  
}\ee
where
\be
  H_{\mu\nu\alpha} \equiv \partial_\m B_{\n\a} + \partial_\n B_{\a\m} + \partial_\a B_{\m\n}  
\ee
The transformation laws that are a symmetry of the Lagrangian~in Equation (\ref{eq:TMLagrangian}) are
\be
 \eqalign{
{\rm D}_a \varphi ~&=~ \chi_a  \cr
{\rm D}_a B{}_{\mu \, \nu } ~&=~ -\, \fracm 14 ( [\, \gamma_{\mu}
\, , \,  \gamma_{\nu} \,]){}_a{}^b \, \chi_b  \cr
{\rm D}_a \chi_b ~&=~ i\, (\gamma^\mu){}_{a \,b}\,  \partial_\mu \varphi 
~-~  (\gamma^5\gamma^\mu){}_{a \,b} \, \e{}_{\mu}{}^{\r \, \s \, \t}
\partial_\r B {}_{\s \, \t}
}  \label{BV5}
\ee

Choosing temporal gauge $B_{0\mu} =0$ and reducing to the 0-brane with nodal field definitions
\begin{subequations}
\label{e:TMnodes}
\begin{align}
	\label{e:TMfermions}
&  i\Psi_1 = \chi_1,~~~i\Psi_2 = \chi_2,~~~i\Psi_3 = \chi_3,~~~i\Psi_4 = \chi_4~~~ \\
	\label{e:TMbosons}
	&\Phi_1 = \varphi ,~~~\Phi_2 = 2 B_{12},~~~\Phi_3 = 2 B_{23},~~~\Phi_4 = 2 B_{31}~~~ 
\end{align}
\end{subequations}
 reduces the Lagrangian~in Equation (\ref{eq:TMLagrangian}) to Equation~(\ref{e:L0}) and transformation laws to Equation~(\ref{e:D}) with the $\brL_\rI$ matrices given by
\begin{align}\label{e:TM}
\mbox{
{
$	\brL_1^{(\text{TM})} =\left(
\begin{array}{cccc}
 1 & 0 & 0 & 0 \\
 0 & 0 & -1 & 0 \\
 0 & 0 & 0 & -1 \\
 0 & -1 & 0 & 0 \\
\end{array}
\right)
$}}
&,~~~
\mbox{
{
$
\brL_2^{(\text{TM})} = \left(
\begin{array}{cccc}
 0 & 1 & 0 & 0 \\
 0 & 0 & 0 & 1 \\
 0 & 0 & -1 & 0 \\
 1 & 0 & 0 & 0 \\
\end{array}
\right)
$}}
,\cr
\mbox{
{
$
\brL_3^{(\text{TM})} = \left(
\begin{array}{cccc}
 0 & 0 & 1 & 0 \\
 1 & 0 & 0 & 0 \\
 0 & 1 & 0 & 0 \\
 0 & 0 & 0 & -1 \\
\end{array}
\right)
$}}
&,~~~
\mbox{
{
$
\brL_4^{(\text{TM})} =\left(
\begin{array}{cccc}
 0 & 0 & 0 & 1 \\
 0 & -1 & 0 & 0 \\
 1 & 0 & 0 & 0 \\
 0 & 0 & 1 & 0 \\
\end{array}
\right)
$}}
\end{align}
and $\brR_\rI$ given by Equation~(\ref{e:adinkraic}) and satisfy the $\mathcal{GR}(4,4)$ algebra in Equation~(\ref{e:GRdN}). By the rules explained in Section~\ref{s:AdinkraReview}, it can be seen that the 0-brane transformation laws for the TM, Equation~(\ref{e:D})  with 
{$\brL_\rI$ and $\brR_\rI$ matrices as in Equations~(\ref{e:TM}) and~(\ref{e:adinkraic})}
and nodal field definitions~(Equation \eqref{e:TMnodes}), are described by the adinkra in Figure~\ref{f:TMAdinkra}. The TM is in the trans-equivalence class ($\chi_0 =-1$) as can be seen through either Equation~(\ref{e:chromo}) or through field redefinitions transforming Figure~\ref{f:TMAdinkra} into the trans-adinkra in Figure~\ref{f:ctAdinkra}.

\begin{figure}[!htbp]
	\centering
	\includegraphics[width = 0.48 \textwidth]{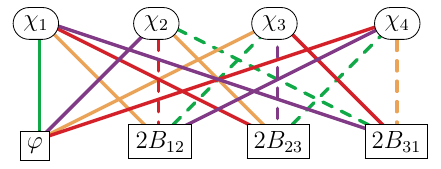}
	\caption{An adinkra for the 0-brane reduced transformation laws of the 4D, $\mathcal{N}=1$ tensor multiplet.}
	\label{f:TMAdinkra}
\end{figure}

\subsection{The 4D, \texorpdfstring{$\mathcal{N}=1$}{N=1} Off-Shell Vector Multiplet (VM)}
The dynamical field content of the VM is a $U(1)$ gauge vector field $A_\mu$ and a Majorana fermion $\lambda_a$. The only auxiliary field in the VM is a pseudoscalar ${\rm d}$. The 4D component Lagrangian for the VM is given by
\be\label{eq:VMLagrangian}\eqalign{
   {\mathcal L}_{\text{VM}} = &-\frac{1}{4}F_{\mu\nu}F^{\mu\nu} +\frac{1}{2}i(\gamma^{\mu})^{ab}\lambda_{a}\partial_{\mu}\lambda_{b}+\frac{1}{2}{\rm d}^2
}\ee 
where
\begin{align}
	F_{\m\n} = \partial_\m A_\n - \partial_\n A_\m
\end{align}

The transformation laws that are a symmetry of the Lagrangian~in Equation (\ref{eq:VMLagrangian}) are
\be  \eqalign{
{\rm D}{}_a A_{\mu} & ~=~ (\gamma_{\mu})_a{}^b \lambda_b  
 ~~~~~~~~~~~~~~~~~~~~~~~~~~~~~\, \cr
{\rm D}{}_a \lambda_b & ~=~ - i \, \fracm 12 \, (  \gamma^{\mu}
\gamma^{\nu} )_{ab} \,  F{}_{\mu \, \nu}  \, ~+~ (\gamma^5
)_{ab} \,   {\rm d}  \cr  
{\rm D}_a {\rm d} & ~=~  i (\gamma^5 \gamma^{\mu})_a{}^b \, \pa_{\mu} \lambda_b 
~~~~~~~~~~~~~~\,~~~~~~~\, 
}  \label{BV3}
\ee
Choosing temporal gauge $A_{0} =0$ and reducing to the 0-brane with nodal field definitions
\begin{subequations}
\label{e:VMnodes}
\begin{align}
	\label{e:VMfermions}
	 &  i\Psi_1 = \lambda_1,~~~i\Psi_2 = \lambda_2,~~~i\Psi_3 = \lambda_3,~~~i\Psi_4 = \lambda_4~~~ \\
	\label{e:VMbosons}
	&\Phi_1 = A_1,~~~\Phi_2 = A_2,~~~\Phi_3 = A_3,~~~\Phi_4 = \int d\tau~{\rm d}~~~ 
\end{align}
\end{subequations}
 reduces the Lagrangian~in Equation (\ref{eq:VMLagrangian}) to Equation~(\ref{e:L0}) and transformation laws to Equation~(\ref{e:D}) with the $\brL_\rI$ matrices given by
\begin{align}\label{e:VM}
\mbox{
{
$	\brL_1^{(\text{VM})} =\left(
\begin{array}{cccc}
 0 & 1 & 0 & 0 \\
 0 & 0 & 0 & -1 \\
 1 & 0 & 0 & 0 \\
 0 & 0 & -1 & 0 \\
\end{array}
\right)
$}}
&,~~~
\mbox{
{
$
\brL_2^{(\text{VM})} =\left(
\begin{array}{cccc}
 1 & 0 & 0 & 0 \\
 0 & 0 & 1 & 0 \\
 0 & -1 & 0 & 0 \\
 0 & 0 & 0 & -1 \\
\end{array}
\right)
$}}
,\cr
\mbox{
{
$
\brL_3^{(\text{VM})} =\left(
\begin{array}{cccc}
 0 & 0 & 0 & 1 \\
 0 & 1 & 0 & 0 \\
 0 & 0 & 1 & 0 \\
 1 & 0 & 0 & 0 \\
\end{array}
\right)
$}}
&,~~~
\mbox{
{
$
\brL_4^{(\text{VM})} =\left(
\begin{array}{cccc}
 0 & 0 & 1 & 0 \\
 -1 & 0 & 0 & 0 \\
 0 & 0 & 0 & -1 \\
 0 & 1 & 0 & 0 \\
\end{array}
\right)
$}}
\end{align}
and $\brR_\rI$ given by Equation~(\ref{e:adinkraic}) and satisfy the $\mathcal{GR}(4,4)$ algebra in Equation~(\ref{e:GRdN}). By the rules explained in Section~\ref{s:AdinkraReview}, it can be seen that the 0-brane transformation laws for the VM, Equation~(\ref{e:D}),  with 
{$\brL_\rI$ and $\brR_\rI$ matrices as in Equations~(\ref{e:VM}) and~(\ref{e:adinkraic})}
and nodal field definitions~(Equation \eqref{e:VMnodes}), are described the adinkra in Figure~\ref{f:VMAdinkra}. The VM is in the trans-equivalence class ($\chi_0 =-1$) as can be seen through either Equation~(\ref{e:chromo}) or through field redefinitions transforming Figure~\ref{f:VMAdinkra} into the trans-adinkra in Figure~\ref{f:ctAdinkra}.

\begin{figure}[!htbp]
	\centering
	\includegraphics[width = 0.48 \textwidth]{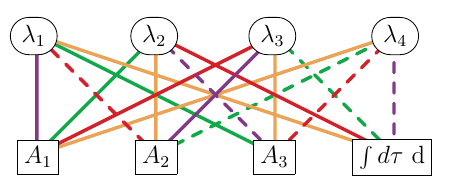}
	\caption{An adinkra for the 0-brane reduced transformation laws of the 4D, $\mathcal{N}=1$ vector multiplet.}
	\label{f:VMAdinkra}
\end{figure}

\subsection{The 4D, \texorpdfstring{$\mathcal{N}=1$}{N=1} Off-Shell Complex Linear Superfield Multiplet (CLS)}\label{s:CLS}
The dynamical field content of the CLS is the same as for the CM: a scalar field and pseudoscalar field, here called $K$ and $L$, respectively, and a Majorana fermion, here called $\zeta_a$. A key difference from CLS and CM is the auxiliary field content: a scalar $M$, pseudoscalar $N$, vector $V_\mu$, pseudovector $U_\mu$, a~low-dimensional fermion $\rho_a$ with $[\rho_a] =3/2$ and a high-dimensional fermion $\beta_a$ with $[\beta_a] = 5/2$. The~4D component Lagrangian for CLS is
\be\label{eq:CLSLagrangian}\eqalign{
\mathcal{L}_{\text{CLS}} &= -\frac{1}{2}\partial_\mu K \partial^\mu K - \frac{1}{2} \partial_\mu L \partial^\mu L - \frac{1}{2} M^2 - \frac{1}{2} N^2 +\frac{1}{4} U_\mu U^\mu  +\frac{1}{4} V_\mu V^\mu \cr
& +\frac{1}{2}i (\gamma^\mu)^{ab} \zeta_a \partial_\mu \zeta_b  + i \rho_a C^{ab}\beta_b 
}\ee
The transformation laws that are a symmetry of the Lagrangian~in Equation (\ref{eq:CLSLagrangian}) are
\begin{align}\label{eq:DCLM1}
\begin{split}
 {\rm D}_a K=& \rho_a - \zeta_a \cr
 {\rm D}_a M =&  \b_a -\frac{1}{2} (\g^\nu)_{a}^{\,\,\,d} \partial_\nu \r_d  \cr
 {\rm D}_a N =&   -i   (\g^5)_a^{\,\,\,\,d} \b_d + \frac{i}{2} (\g^5 \g^\nu)_{a}^{\,\,\,d} \partial_\nu  \r_d \cr
 {\rm D}_a L =&   i (\g^5)_a^{\,\,\,d} ( \r_d + \z_d) \cr
 {\rm D}_a U_\mu = & i (\g^5 \g_\mu)_a^{\,\,\,\,d} \b_d - i (\g^5)_a^{\,\,\,\,d}   \partial_\mu  (\r_d + 2\z_d) -  \frac{i}{2}\, (\g^5 \g^\nu \g_\mu)_{a}^{\,\,\,d} \partial_\nu(\r_d - 2\z_d) \cr
{\rm D}_a V_\mu = & - (\g_\mu)_a^{\,\,\,\,d} \b_d +   \partial_\mu (\r_a - 2\z_a) +  \frac{1}{2} ( \g^\nu \g_\mu)_{a}^{\,\,\,d} \partial_\nu (\r_d + 2\z_d) \cr
{\rm D}_a \z_b = & - i  (\g^\mu)_{ab} \partial_\mu K - (\g^5 \g^\mu)_{ab} \partial_\mu L
-  \, \frac{1}{2}( \g^5 \g^\mu)_{ab} U_\mu + i \frac{1}{2} (\g^\mu)_{ab}  V_\mu   \cr
 {\rm D}_a \r_b = &    i C_{ab} M  + (\g^5)_{a b} N + \,\frac{1}{2}( \g^5 \g^\mu)_{ab} U_\mu +  i \frac{1}{2} (\g^\mu)_{ab}  V_\mu \cr
 {\rm D}_a \b_b = & \frac{i}{2} (\g^\mu)_{ab} \partial_\mu M + \frac{1}{2} (\g^5 \g^\mu)_{ab} \partial_\mu N  + \frac{1}{2} (\g^5 \g^\mu \g^\nu)_{ab} \partial_\mu U_\nu \cr
& + \frac{1}{4} (\g^5 \g^\nu \g^\mu )_{ab}\partial_\mu U_\nu + \frac{i}{2} (\g^\mu \g^\nu)_{ab} \partial_\mu V_\nu + \frac{i}{4} \,(\g^\nu \g^\mu)_{ab} \partial_\mu V_\nu 
\cr
& + \eta^{\mu\nu}\partial_\mu \partial_\nu (- i C_{ab} K +(\g^5)_{ab} L ).
\end{split}
\end{align}

Unlike the minimal CM, TM, and VM cases, the 0-brane reduction for CLS requires nodal definitions that are linear combinations of the 0-brane reduced fields for the resulting $\brL_{\rm I}$ matrices to be adinkraic (Equation~(\ref{e:adinkraic})).  A particular choice of nodal field definitions for CLS that are adinkraic are
\be\label{e:CLSnodes}
  \dot{\Phi} =
\left(
\begin{array}{c}
 -M \\
 \dot{K}-V_0 \\
 -\dot{L}-U_0 \\
 N \\
 U_2 \\
 V_0-2 \dot{K} \\
 -U_1 \\
 U_3 \\
 -V_3 \\
 V_1 \\
 -2 \dot{L}-U_0 \\
 V_2
\end{array}
\right)
,~~~i \dot{\Psi} = \left(
\begin{array}{c}
 \frac{\dot{\r}_2}{2}-\beta _1 \\
 -\beta _2-\frac{\dot{\r}_1}{2} \\
 -\beta _3-\frac{\dot{\r}_4}{2} \\
 \beta _4-\frac{\dot{\r}_3}{2} \\
 \beta _1-\dot{\z}_2+\frac{\dot{\r}_2}{2} \\
 \beta _2+\dot{\z}_1-\frac{\dot{\r}_1}{2} \\
 \beta _3+\dot{\z}_4-\frac{\dot{\r}_4}{2} \\
 \beta _4-\dot{\z}_3+\frac{\dot{\r}_3}{2} \\
 \beta _1+\dot{\z}_2+\frac{\dot{\r}_2}{2} \\
 -\beta _2+\dot{\z}_1+\frac{\dot{\r}_1}{2} \\
 -\beta _3+\dot{\z}_4+\frac{\dot{\r}_4}{2} \\
 \beta _4+\dot{\z}_3+\frac{\dot{\r}_3}{2}
\end{array}
\right)
\ee
Here, we   define  $\Phi$ and $\Psi$ in terms of their derivatives $\dot{\Phi}$ and $\dot{\Psi}$ for notational convenience~(integration constants are assumed to be zero). 
With these nodal field definitions, the~Lagrangian~in Equation (\ref{eq:CLSLagrangian}) reduces to Equation~(\ref{e:L0}) and transformation laws to Equation~(\ref{e:D}) with the $\brL_\rI$ matrices given by
\begin{align}
\label{e:CLS}
\begin{split}
\mbox{{
${\bm {\rm L}}_1^{(\text{CLS})} =  {\bm {\rm  I}}_3  \otimes {\bm {\rm I}}_4
$}}
  ,~~~&
  \mbox{{ 
${\bm {\rm L}}_2^{(\text{CLS})} = i {\bm {\rm I}}_3 \otimes {\bm \beta}_3 
$}}
  \\
  \mbox{{ 
${\bm {\rm L}}_3^{(\text{CLS})} = i \left(\begin{array}{ccc}
                        1 & 0 & 0 \\
                        0 & -1 & 0 \\
                        0 & 0 & -1
                   \end{array}\right) \otimes  {\bm \beta}_2
                   $}}
,~~~&
\mbox{{ 
${\bm {\rm L}}_4^{(\text{CLS})} = - i {\bm {\rm I}}_3 \otimes {\bm \beta}_1 
$}}
\end{split}
\end{align}
with the 
${\bm \alpha}$ and ${\bm \beta}$ 
matrices as in Appendix~\ref{a:ab} and $\brR_\rI$ given by Equation~(\ref{e:adinkraic}).

By the rules explained in Section~\ref{s:AdinkraReview}, it can be seen that the 0-brane transformation laws for the CLS, Equation~(\ref{e:D})  with $\brL_\rI$ matrices as in Equation~(\ref{e:CLS}) and nodal field definitions~(Equation \eqref{e:CLSnodes}), are described by the adinkra in Figure~\ref{f:CLSAdinkra}. The CLS has $\chi_0 =-1$ as can be seen through either Equation~(\ref{e:chromo}) or through the fact that comparing Figure~\ref{f:CLSAdinkra} with Figure~\ref{f:ctAdinkra}, Figure~\ref{f:CLSAdinkra} is one cis-adinkra ($n_c = 1$) and two trans-adinkras ($n_t=2$), thus $\chi_0 = n_c - n_t = -1$. 

\begin{figure}[!htbp]
	\centering
	\includegraphics[width = 0.9\textwidth]{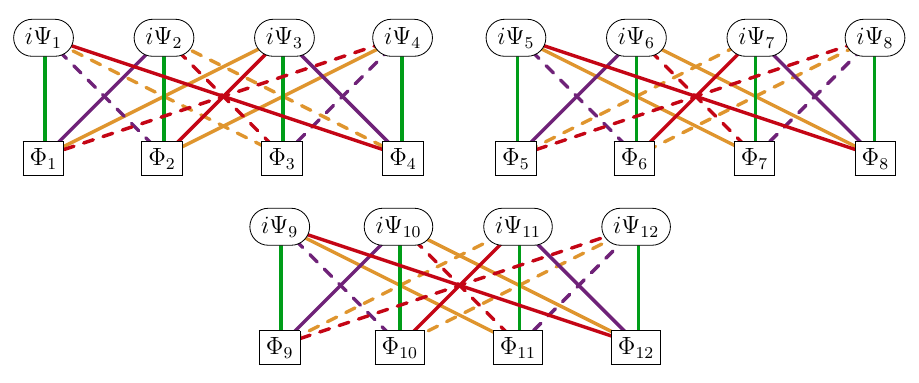}
	\caption{An adinkra for the 0-brane reduced transformation laws of the 4D, $\mathcal{N}=1$ complex linear superfield multiplet (CLS) and old-minimal supergravity multiplet (mSG).}
	\label{f:CLSAdinkra}
\end{figure}

\subsection{The 4D, \texorpdfstring{$\mathcal{N}=1$}{N=1} Off-Shell Old-Minimal Supergravity Multiplet (mSG)}\label{s:mSG}
The dynamical field content of mSG is the symmetric, rank two graviton $h_{\mu\nu}$ and gravitino $\psi_{\mu a}$.  The auxiliary fields of mSG are a scalar $S$, pseudoscalar $P$, and axial vector $A_\mu$. The 4D component Lagrangian for mSG is
\begin{align}\label{eq:mSGLagrangian}
     \mathcal{L}_{mSG} = & -\frac{1}{2}\partial_\alpha h_{\mu\nu} \partial^\alpha h^{\mu\nu} + \frac{1}{2}\partial^\alpha h \partial_\alpha h - \partial^\alpha h \partial^\beta h_{\alpha\beta} + \partial^\mu h_{\mu\nu} \partial_\alpha h^{\alpha\nu}  \cr
     &- \frac{1}{3}S^2 -\frac{1}{3}P^2 + \frac{1}{3} A_\mu A^\mu -\frac{1}{2} \psi_{\mu a} \epsilon^{\mu\nu\alpha\beta}(\gamma^5\gamma_\nu)^{ab} \partial_\alpha \psi_{\beta b}
\end{align}
The transformation laws that are a symmetry of the Lagrangian~in Equation (\ref{eq:mSGLagrangian}) are
\begin{subequations}\label{eq:Dfinalnumeric}
\begin{align}
   {\rm D}_a S =~&  -\frac{1}{2} ([\g^\m , \g^\n ])_{a}^{~b}\partial_\mu \psi_{\nu b} \\
   {\rm D}_a P =~& \frac{1}{2}(\gamma^5[ \g^\m , \g^\n])_{a}^{~b}\partial_\mu \psi_{\nu b} \\
   {\rm D}_a A_{\mu} =~& i (\gamma^5 \gamma^\nu)_a^{~b} \partial_{[\nu} \psi_{\mu] b} -\frac{1}{2} \epsilon_{\mu}^{~\nu\alpha\beta}(\gamma_\nu)_{a}^{~b} \partial_\alpha \psi_{\beta b}  \\
   {\rm D}_a h_{\mu\nu} =~& \frac{1}{2} (\gamma_{(\mu})_{a}^{~b}\psi_{\nu) b} \\
   {\rm D}_a \psi_{\mu b} =~&  - \frac{i}{3} (\gamma_\mu)_{ab} S -\frac{1}{3} (\gamma^5\gamma_\mu)_{ab} P + \frac{2}{3} (\gamma^5)_{ab} A_\mu +\frac{1}{6} (\gamma^5[\g_{\mu}, \g^{\nu}])_{ab}A_\nu  + \cr
   & - \frac{i}{2}([\g^{\alpha} , \g^{\beta}])_{ab}\partial_\alpha h_{\beta\mu} 
\end{align}
\end{subequations}

Similar to the CLS, the 0-brane reduction for mSG requires nodal definitions that are linear combinations of the 0-brane reduced fields for the resulting $\brL_{\rm I}$ matrices to be adinkraic (Equation~(\ref{e:adinkraic})).  As shown in~\cite{Chappell:2012qf}, the most general choice of nodal field definitions for mSG that are adinkraic and match with the cis- and/or trans-adinkra Figure~\ref{f:ctAdinkra} are
 \begin{equation}\label{e:mSGnodes}
\Phi  =  \left(
\begin{array}{c}
A_0 \\
P \\
-S \\
\dot{h}\\
\hline
\vspace{-7 pt} \\
 u_1 A_1
   +(u_2-u_3 )\dot{h}_{23} \\
 u_3 A_3 +
   (u_1-u_2)\dot{h}_{12} \\
 u_2 A_2 +(u_3-u_1 )\dot{h}_{31}
   \\
-u_1\dot{h}_{11} -u_2\dot{h}_{22} -u_3\dot{h}_{33}
   \\
   \hline
\vspace{-7 pt} \\
v_1 A_1
   +(v_2-v_3 )\dot{h}_{23}  \\
 v_3 A_3 +
   (v_1-v_2)\dot{h}_{12} \\
 v_2 A_2 +(v_3-v_1 )\dot{h}_{31}
   \\
-v_1\dot{h}_{11} -v_2\dot{h}_{22} -v_3\dot{h}_{33}
\end{array}
\right),
~~~i \Psi = \left(
\begin{array}{c}
\dot{\psi} _{13}-\dot{\psi} _{21}-\dot{\psi} _{34} \\
-\dot{\psi} _{14}-\dot{\psi} _{22}-\dot{\psi} _{33} \\
 \dot{\psi} _{11}+\dot{\psi} _{23}-\dot{\psi} _{32} \\
 \dot{\psi} _{12}-\dot{\psi} _{24}+\dot{\psi} _{31} \\
\hline
\vspace{-7 pt} \\
 -u_1 \dot{\psi }_{13}+u_2
   \dot{\psi }_{21}+u_3 \dot{\psi }_{34} \\
-u_1 \dot{\psi }_{14}-u_2 \dot{\psi }_{22}-u_3 \dot{\psi
   }_{33}\\
-u_1 \dot{\psi }_{11}-u_2 \dot{\psi }_{23}+u_3 \dot{\psi
   }_{32}\\
 -u_1 \dot{\psi }_{12}+u_2
   \dot{\psi }_{24}-u_3 \dot{\psi }_{31}\\
 \hline
\vspace{-7 pt} \\
 -v_1 \dot{\psi }_{13}+v_2
   \dot{\psi }_{21}+v_3 \dot{\psi }_{34} \\
-v_1 \dot{\psi }_{14}-v_2 \dot{\psi }_{22}-v_3 \dot{\psi
   }_{33} \\
-v_1 \dot{\psi }_{11}-v_2 \dot{\psi }_{23}+v_3 \dot{\psi
   }_{32} \\
 -v_1 \dot{\psi }_{12}+v_2
   \dot{\psi }_{24}-v_3 \dot{\psi }_{31}
\end{array}
\right)
\end{equation}
with
\begin{align}\label{e:uconstraint}
   u_1 + u_2 + u_3 = v_1 + v_2 + v_3 = 0
\end{align} 

Here, we   define  $\Phi$ and $\Psi$ in terms of their derivatives $\dot{\Phi}$ and $\dot{\Psi}$ for notational convenience as in the CLS case (integration constants are assumed to be zero). 
With these nodal field definitions, the~Lagrangian~in Equation (\ref{eq:mSGLagrangian}) reduces to Equation~(\ref{e:L0}) and transformation laws to Equation~(\ref{e:D}) with the $\brL_\rI$ matrices given by
\begin{align}
\label{e:mSG}
\begin{split}
\mbox{{
${\bm {\rm L}}_1^{(\text{mSG})} =  {\bm {\rm  I}}_3  \otimes {\bm {\rm I}}_4
$}}
 ,~~~&
  \mbox{{ 
${\bm {\rm L}}_2^{(\text{mSG})} = i {\bm {\rm I}}_3 \otimes {\bm \beta}_3 
$}}
  \\
  \mbox{{ 
${\bm {\rm L}}_3^{(\text{mSG})} = i \left(\begin{array}{ccc}
                        1 & 0 & 0 \\
                        0 & -1 & 0 \\
                        0 & 0 & -1
                   \end{array}\right) \otimes  {\bm \beta}_2
                   $}}
,~~~&
\mbox{{ 
${\bm {\rm L}}_4^{(\text{mSG})} = - i {\bm {\rm I}}_3 \otimes {\bm \beta}_1 
$}}
\end{split}
\end{align}
with the 
${\bm \alpha}$ and ${\bm \beta}$ 
matrices as in Appendix~\ref{a:ab} and $\brR_\rI$ given by Equation~(\ref{e:adinkraic}).

By the rules explained in Section~\ref{s:AdinkraReview}, it can be seen that the 0-brane transformation laws for the mSG, Equation~(\ref{e:D})  with $\brL_\rI$ matrices as in Equation~(\ref{e:mSG}) and nodal field definitions~(Equation \eqref{e:mSGnodes}), are described by the adinkra in Figure~\ref{f:CLSAdinkra}: the same adinkra as for CLS. Thus, as in the CLS case, the mSG has $\chi_0 =-1$ and can be decomposed as one cis-adinkra ($n_c = 1$) and two trans-adinkras ($n_t=2$) with $\chi_0 = n_c - n_t = -1$. 

\subsection{Gadgets}\label{s:GadgetReview}
There is not a tremendous amount of diversity in the $\chi_0$ values reviewed thus far, as summarized in Table~\ref{t:genomicssummary}: all those reviewed in the previous five sections have $\chi_0 = -1$ aside from the CM, which has $\chi_0=+1$. Clearly, another tool is needed to further separate out at the adinkra level which adinkras relate to which higher dimensional systems. 

The \emph{gadget} $\mathcal{G}(\mathcal{R}, \mathcal{R}')$ between two different adinkra representations $\mathcal{R}$ and $\mathcal{R}'$ of the $GR(\rd,N)$ algebra, defined below, is used to separate multiplets at the adinkra level that holographically correspond to different multiplets in higher dimensions~\cite{KIAS1,KIAS2,H3,H2,Gates:2017tly,Gates:2017eui}. 
\begin{align}\label{e:Gadget}
\mathcal{G}(\mathcal{R}, \mathcal{R}') ~&=~  \frac{(N-2)!}{4 (N!)} \sum_{I,J} 
\brtV_{\rm IJ}^{(\mathcal{R})}  \brtV_{\rm IJ}^{(\mathcal{R}')} ~=~   \frac{(N-2)!}
{{\rm d}_{min}(4) (N!)} \sum_{I,J} \brtV_{\rm IJ}^{(\mathcal{R})}  \brtV_{\rm IJ}^{
(\mathcal{R}')} 
\end{align}

In the above, the function ${\rm d}_{min}(4)=4$ is the minimal size of an $N=4$ adinkra as proved for general $N$ in~\cite{Doran:2008kg,Doran:2011gb} and used in the subsequent works~\cite{Doran:2013vea,Gates:2017tly}. 
{
For 4D, $\mathcal{N} =1$ supersymmetry, 
the number of colors in the adinkraic representation is $N=4$.}

{The gadget between two representations is analogous to a \emph{dot product} between two vectors. Two representations that have the same holoraumy $\brtV_{\rm IJ}$  are known as $\brtV_{\rm IJ}$-equivalent and will have a gadget value of $n_c + n_t$. Owing to the dot product analogy, representations that are $\brtV_{\rm IJ}$-equivalent  are analogous to parallel vectors. Representations that have gadget value different from $n_c + n_t$ are said to be $\brtV_{\rm IJ}$-inequivalent and are analogous to vectors at an angle to one another. An interesting case is therefore when two representations have a gadget of zero: we term such representations \emph{gadget-orthogonal} in the analogous sense to orthogonal vectors. As such, gadget-orthogonal representations are considered the most distinct two representations can be, in the sense of holoraumy.   
}

Gadgets can only be compared between systems of the same size $d$ and number of colors $N$. The~CM, TM, and VM all have $d=4$, $N=4$, thus the gadget may be calculated amongst them. As first discovered in~\cite{KIAS1,KIAS2}, the gadgets between the representations ordered $\mathcal{R} = \mathcal{R'} = (CM, TM, VM)$ take the following form 
\begin{align}\label{e:Gmin}
	\mathcal{G}(\mathcal{R}, \mathcal{R}') = \begin{pmatrix}
		1 & 0 & 0 \\
		0 & 1 & -1/3\\
		0 & -1/3 & 1
	\end{pmatrix}
\end{align}


The result in Equation~(\ref{e:Gmin}) demonstrates that the CM is gadget-orthogonal to the TM and VM and thus can be thought of as distinct at the adinkra level. The cis  and trans  content already demonstrate  this: the CM has $\chi_0 = +1$ ($n_c = 1,n_t = 0$) and both the VM and TM have $\chi_0 = -1$  ($n_c=0,n_t = 1$). More importantly, Equation~(\ref{e:Gmin}) demonstrates that the 
VM and TM are $\tilde{V}_{\rm IJ}$-inequivalent, thus \emph{can be thought of as distinct even at the adinkra level.} 
Thus, the gadget separates the TM and VM at the adinkra level, and so is a further distinguishing calculation that can be done in addition to $\chi_0$. 

The gadget is invariant with respect to some nodal field redefinitions and not invariant with respect to others: this depends on whether an adinkra's holoraumy changes     under the redefinition, as investigated in~\cite{Gates:2017eui}. No bosonic field redefinition can change the gadget as defined in Equation~(\ref{e:Gadget}) as the $\brtV_{\rm IJ}$ matrices can only act on fermions from either the left or the right: the bosonic indices are fully contracted in the $\brR_\rI\brL_\rJ$ multiplication similar to how a Lorentz scalar's spacetime indices are fully contracted. As such, we   concern ourselves with fermionic field redefinitions in this paper, and the diagonal Lagrangian parameter we   see in the $20 \times 20$ transformation laws   pertains only to a fermionic field redefinition symmetry of the Lagrangian.

In constructing a Dykin diagram, it is necessary to choose a convention as to which roots to use: the simple roots are the canonical choice~\cite{Georgi:1999wka}. Similarly, it is necessary to follow a nodal field definition convention in comparing gadget values. The CM, TM, and VM adinkras are defined according to the following convention. Define the fermion nodes such that the node number matches the component number. For bosons, place them in the nodes left to right with dynamical fields first and auxiliary fields second. Within the lists of dynamical and auxiliary fields, place them left to right as follows: scalars, pseudoscalars, vectors, pseudovectors, tensors, and pseudotensors. List vectors in numerical order, as in the VM and list tensors in the order as demonstrated for the TM: components 12, 23, and 31. For the $20 \times 20$ multiplets investigated in this paper, we   expand these rules to be applicable to larger multiplets.

The one-to-one nodal field definitions for the CM, VM, and TM used in Equations~(\ref{e:CMnodes}),~(\ref{e:TMnodes}), and~(\ref{e:VMnodes}) result  in adinkraic $\brL_\rI$ and $\brR_\rI$ matrices that satisfy the relationship in Equation~(\ref{e:adinkraic}). As~such, these multiplets can     be expressed as the adinkras in Figures~\ref{f:CMAdinkra}, \ref{f:TMAdinkra}, and~\ref{f:VMAdinkra} with single fields corresponding to each node. No such one-to-one nodal field definitions are possible for the mSG and CLS as the nodes of the adinkra must necessarily correspond to linear combinations of the fields, as shown in Equations~(\ref{e:CLSnodes}) and~(\ref{e:mSGnodes}). We show this is the case for the $20 \times 20$ multiplets investigated in this paper: one-to-one nodal field definitions do not lead to adinkraic $\brL_\rI$ and $\brR_\rI$ matrices. On a final note, the value of $\chi_0$ is insensitive to basis choice, owing to the trace over which it is defined and the fact that it is defined only over a single representation as shown in Equation~(\ref{e:chromo}).

\newpage

\section{The de Wit--van Holten Formulation}\label{s:dWvH}
We refer to the matter gravitino  multiplet as described in Refs.\ \cite{dWvH,Fradkin:1979as,MGM} as the ``de Wit--van Holten'' (dWvH)
formulation (the labeling of this multiplet as dWvH is due to the fact that it appeared as a 4D, $\cal N$ = 1 submultiplet \cite{dWvH,Fradkin:1979as} prior to the work of \cite{MGM}). 
The dWvH multiplet consists of a spin 
one-half superfield with compensators of a vector multiplet and chiral multiplet~\cite{Gates:1983nr}. 
The~components of this multiplet are as follows. The~matter fields are that of a spin 
three-halves Rarita Schwinger field $\psi_{\mu b}$ and a spin one vector $B_{
\mu}$. The~bosonic auxiliary fields in the multiplet (all with dimension-two) are a 
scalar $K$, pseudoscalars $L$ and $P$, rank-two tensor $t_{\mu\nu}$, vector $
V_\mu$, and axial vector $U_{\mu}$. The fermionic auxiliary fields are a dimension 
three-halves spinor $\lambda_a$ and dimension five-halves spinor $\chi_a$. The~
transformation laws, Lagrangian, algebra, and adinkras are described in the 
following subsections in a real Majorana~notation.

\subsection{Transformation Laws}
We write the transformation laws in terms of a single free parameter $c_0$, which parameterizes a field redefinition of the fermionic fields that leaves the Lagrangian invariant.
\begin{align}
{\rm D}_a \lambda_b & \,= i \frac{1}{2} C_{ab} K + \frac{1}{2} (\g^5)_{ab} L +  
(\g^5)_{ab} P - i \frac{1}{8} ([\g^\mu,\, \g^\nu])_{ab} t_{\mu \nu} - 
i \frac{1}{2} (\g^\mu)_{ab} V_\mu +\frac{1}{2} (\g^5 \g^\mu)_{ab} 
U_\mu \\[6pt]
{\rm D}_a \chi_b & \, = -i \frac{1}{2} c_1 (\g^\mu)_{ab} \partial_\mu K -\frac{1}{2} c_1 (\g^5 \g^\mu)_{ab} \partial_\mu L -c_3 (\g^5 \g^\mu)_{ab} \partial_\mu P \cr
& \hspace{15pt} + i \frac{1}{2} c_2  (\g^\mu \g^\nu)_{ab} \partial_\nu V_\mu +  i c_0 (\g^\mu \g^\nu)_{ab} \partial_\mu V_\nu  -\frac{1}{2} c_2 (\g^5\g^\mu\g^\nu)_{ab}\partial_\nu U_\mu  - c_0 (\g^5\g^\mu\g^\nu)_{ab} \partial_\mu U_\nu \cr
& \hspace{15pt} 
+ i \frac{1}{8}c_3([\g^\mu , \g^\nu]\g^\alpha)_{ab} \partial_\alpha 
t_{\mu \nu} + i\frac{1}{4}c_4 (\g^\alpha[\g^\mu,\g^\nu]_{ab} \partial_\alpha W_{\mu\nu}\\[6pt]
{\rm D}_a \psi_{\mu b} & \, =  i \frac{1}{2} c_4 (\g_\mu)_{ab} K + \frac{1}{2}c_4 (\g^5\g_\mu)_{ab} L  +c_0 (\g^5 \g_\mu)_{ab} 
P + i \frac{1}{2} c_4 (\g_\mu\g^\nu)_{ab} V_\nu -i c_0 C_{ab} V_\mu \cr
&\hspace{15 pt} - \frac{1}{2} c_4 (\g^5\g_\mu\g^\nu)_{ab} U_\nu +c_0 (\g^5)_{ab} U_\mu + i \frac{1}{8} (\g_\mu [\g^\alpha, \g^\beta])_{ab} W_{\alpha\beta} - i \frac{1}{8} c_0 ([\g^\alpha, \g^\beta]\g_\mu)_{ab} t_{\alpha\beta} \\[6pt]
{\rm D}_a P & \, = - i \frac{1}{2} (\g^5)_a^{~b}(\chi_b + c_0 R_b)  - i \frac{1}{2} c_3 (\g^5\g^\mu)_a^{~b}\partial_\mu \lambda_b \\[6pt]
{\rm D}_a K & \, =  \frac{1}{2}(\chi_a + c_4 R_a) + \frac{1}{2}c_1 (\g^\mu)_a{}^b \partial_\mu \lambda_b  
\\[6pt]
{\rm D}_a L & \, = -i (\g^5)_a^{~b} ({\rm D}_b K) \\[6pt]
{\rm D}_a V_\mu & \, =  \frac{1}{2}(\g_\mu)_a{}^b (\chi_b + c_0 R_b) +\frac{1}{2} c_2 (\g_\mu \g^\nu)_a{}^b \partial_\nu \lambda_b + c_0(\g^\nu\g_\mu )_a{}^b \partial_\nu 
\lambda_b + (\g^\nu)_{a}{}^{b} \partial_{[\mu} \psi_{\nu ] b} \\[6pt]
{\rm D}_a U_\mu & \, = i (\g^5)_a{}^b ({\rm D}_b V_\mu) \qquad  \\[6pt]
{\rm D}_a B_\mu & \, = \psi_{\mu a} + c_4 (\g_{\mu})_a{}^b \lambda_b  \\[6pt]
{\rm D}_a t_{\mu \nu} & \, = \frac{1}{4} ([\g_\mu,\, \g_\nu])_a{}^b (\chi_b + c_0 R_b) +\frac{1}{2} c_5 (\g_{[\mu})_{|a|}^{~~b}\partial_{\nu]} \lambda_b - i  \frac{1}{2} c_6  \epsilon_{\mu\nu\alpha\beta}(\g^5\g^\alpha)_a^{~b} \partial^\beta \lambda_b \cr
&\hspace{15 pt} + \partial_{[\mu} 
\psi_{\nu] a} + i \epsilon_{\mu\nu}{}^{\alpha\beta}(\g^5)_a{}^b \partial_\alpha \psi_{\beta b} 
\end{align}
\vspace*{8pt}
where
\begin{align}
c_1 &= 3 c_0^2 + 6 c_0 +  1,~~~c_2 = 3 c_0^2 + 4 c_0 + 1,~~~c_3 = 3 c_0^2 - 1,~~~c_4 = c_0 + 1 \\
c_5 &= 3 c_0^2  - 2 c_0 - 3,~~~c_6 = 3 c_0^2 + 2 c_0 + 1.
\end{align}
The gauge-invariant fields strengths are $R_a \equiv ([\g^\mu,\, \g^\nu])_a{}^b \partial_\mu  \psi_{\nu b}$, $F_{\mu \nu} 
\equiv \partial_{[\mu} B_{\nu]}$, and \mbox{$W_{\mu\nu} = t_{\mu\nu} - 2 F_{\mu\nu}$.}

Using Fierz identities, the term including $F_{\alpha\beta}$ within $W_{\alpha\beta}$ of the transformation law for the gravitino can be 
expressed as follows:
\begin{align}
-i\frac{1}{4} (\g_\mu[\g^\alpha , \g^\beta])_{ab} F_{\alpha \beta} =&  - i (\g^\nu)_{ab} \partial_\mu B_\nu + i 
(\g^\nu)_{ab} \partial_\nu B_\mu + \epsilon_{\mu}^{~\nu\alpha\beta}(\g^5\g_\nu)_{ab} 
\partial_\alpha B_\beta.
\end{align}

As the first term encodes the gravitino's gauge transformation, it can be ignored. 
This is true certainly at the adinkraic level, where this term only shows up in the 
transformation laws for $\psi_{0b}  =0$ in temporal~gauge.

\subsection{Anti-Commutators}
Direct calculations of the anti commutators of the D-operators on all the fields yield
the results:
\begin{align}
\{ {\rm D}_a,\, {\rm D}_b \}X \, = & 2 i (\g^{\mu})_{ab} \partial_{\mu} X ~~ \mbox{for} 
~~ X \in \left\{ \lambda_c,\, \chi_c,\, P,\, K,\, L,\, V_\mu,\, U_\mu,\, t_{\mu \nu} \right\} \\[6pt]
\{ {\rm D}_a,\, {\rm D}_b \} B_\mu \, = & 2 i (\g^{\alpha})_{ab} \partial_{\alpha} B_{\mu} 
- \partial_\mu \Lambda_{ab} \\[6pt]
\{ {\rm D}_a,\, {\rm D}_b \} \psi_{\mu c} \, = & 2 i (\g^\alpha)_{ab} \partial_\alpha \psi_{
\mu c} - \partial_\mu \epsilon_{abc}
\end{align}
where
\vspace{-12 pt}
\begin{subequations}\label{e:dWvHNonClosure}
	\begin{align}
	\Lambda_{ab} =& 2 i (\g^\nu)_{ab} B_\nu \\
	\epsilon_{abc} =& 2 i (\g^\nu)_{ab} \psi_{\nu c} + 2 i c_0 (\gamma^\nu)_{ab} (\gamma_\nu)_c{}^d \lambda_d.
	\end{align}
\end{subequations}
The non-closure terms $\Lambda_{ab}$ and $\epsilon_{abc}$ indicate gauge transformations of the $B_\mu$ and $\psi_{\mu a}$ fields. As such, the algebra closes on the field strengths $F_{\mu\nu}$ and $R_a$:
\begin{align}
	\{ {\rm D}_a,\, {\rm D}_b \} F_{\mu\nu} \, = & 2 i (\g^{\alpha})_{ab} \partial_{\alpha} F_{\mu\nu} \cr
	 \{ {\rm D}_a,\, {\rm D}_b \} R_c \, = & 2 i (\g^{\alpha})_{ab} \partial_{\alpha} R_c
\end{align}

\subsection{Lagrangian}
The Lagrangian that is  invariant (up to a surface term) with respect to these transformation
laws takes the form 
\begin{align}
\mathcal{L}  \, = & - P^2 - \frac{1}{2} K^2 - \frac{1}{2} L^2 + \frac{1}{2} V_\mu V^\mu 
+\frac{1}{2} U_\mu U^\mu - \frac{1}{2} F_{\mu\nu}F^{\mu\nu} + \frac{1}{4} t_{\mu\nu}
t^{\mu\nu} \cr
& - \frac 12 \,  \psi_{\mu a} \epsilon^{\mu\nu\alpha\beta} (\g^5 \g_\nu)^{ab} 
\partial_\alpha \psi_{\beta b} + i \lambda_b \chi^b.
\end{align}
The Lagrangian is invariant with respect to the following $c_0$-dependent, fermionic field redefinitions as pointed out in~\cite{MGM}
\begin{align}\label{e:fermredef}
  \psi_{\mu a} \to \psi_{\mu a} + c_0 (\g_\m)_{a}{}^b \lambda_b,~~~\lambda_a \to \lambda_a,~~~\chi_a \to \chi_a + c_0 R_a + 3 c_0^2 (\g^\m)_a{}^b\partial_b \lambda_b
\end{align}
This Lagrangian is also invariant with respect to the following gauge transformations that are indicated by~ Equation~(\ref{e:dWvHNonClosure})
\begin{subequations}\label{e:dWvHgt}
\begin{align}
	\delta B_\mu =~& \partial_\mu \Lambda \\
	\delta \psi_{\mu a} =~& \partial_\mu \epsilon_{a}.
\end{align}
\end{subequations}


\section{The Ogievetsky--Sokatchev (OS) Formulation}\label{s:OS}
The matter gravitino  multiplet, as described in Refs. \cite{OS1,OS2},  consists of a spin one-half 
superfield with compensators of a vector multiplet and tensor multiplet~\cite{Gates:1983nr}. The 
components of this multiplet are as follows. The matter fields are that of a spin 
three-halves Rarita Schwinger field $\psi_{\mu b}$ and a spin one vector $B_{
\mu}$. The bosonic auxiliary fields (all with dimension-two) are a pseudoscalar 
$P$, rank-two tensor $t_{\mu\nu}$,  vector $V_\mu$, axial gauge vector $A_{\mu}$, and 
divergenceless axial vector $G^\mu$ that is actually the field strength of a gauge 
two form $E_{\alpha\beta}$ such that $G^\mu = \frac{1}{2}\epsilon^{\mu\nu\alpha
\beta}\partial_\nu E_{\alpha\beta}$. The fermionic auxiliary fields are a dimension 
three-halves spinor $\xi_a$ and dimension five-halves spinor $\chi_a$. The 
transformation laws, Lagrangian, algebra, and adinkras are described in the 
following subsections in a real Majorana~notation.

\subsection{Transformation Laws}

We write the transformation laws in terms of a single free parameter $s_0$, which parameterizes a field redefinition of the fermionic fields as in Equation~(\ref{e:fermredef}) but with $c_0 \to s_0$ that leaves the Lagrangian~invariant.

\begin{align}
{\rm D}_a \lambda_b & \, =  (\g^5)_{ab} P - i \frac{1}{8} ([\g^\mu,\, \g^\nu])_{ab} 
t_{\mu \nu} - i \frac{1}{2} (\g^\mu)_{ab} V_\mu +\frac{1}{2} (\g^5 \g^\mu)_{ab} G_\mu 
\\[6pt]
{\rm D}_a \chi_b & \, = -s_1 (\g^5 \g^\mu)_{ab} \partial_\mu P + i \frac{1}{2} s_2 C_{ab}\partial_\mu V^\mu - i \frac{1}{4}s_3 [\g^\mu , \g^\nu]_{ab}\partial_\mu V_\nu  - s_0 (\g^5[\g^\mu,\g^\nu])_{ab}\partial_\mu A_\nu \cr
&~~  + \frac{1}{4} s_5 (\g^5 [\g^\mu , \g^\nu])_{ab} \partial_\mu G_\nu + \frac{1}{2} s_1 \epsilon^{\mu\nu\alpha\beta}(\g^5\g_\mu)_{ab}\partial_\nu t_{\alpha\beta}  - i \frac{1}{8}  (\g^\alpha [\g^\mu, \g^\nu])_{ab} \partial_\alpha ( s_6 t_{\mu\nu}+ 4 s_4 F_{\mu\nu}) \\
{\rm D}_a \psi_{\mu b} & \, =  s_0 (\g^5\g_\mu)_{ab}P + i C_{ab} V_\mu - i \frac{1}{2} s_4 (\g^\nu \g_\mu)_{ab} V_\nu - (\g^5)_{ab} A_\mu - \frac{1}{2} (\g^5)_{ab} G_\mu +\frac{1}{2} s_4 (\g^5\g^\nu\g_\mu)_{ab} G_\nu  \cr 
&\hspace{15 pt} + \frac{1}{2}s_0\epsilon_{\mu\nu\alpha\beta}(\g^5\g^\nu)_{ab}t^{\alpha\beta} + i\frac{1}{8} (\g_\mu[\g^\alpha , \g^\beta])_{ab} (s_4 t_{\alpha\beta} - 2 F_{\alpha \beta}) \\[6pt]
{\rm D}_a P & \, = - i \frac{1}{2} (\g^5)_a^{~b}(\chi_b + s_0 R_b) - i \frac{1}{2} s_1 (\g^5\g^\mu)_a^{~b}\partial_\mu \lambda_b \\[6pt]
{\rm D}_a V_\mu & \, =  \frac{1}{2}(\g_\mu)_a{}^b ( \chi_b + s_0R_b) + \frac{1}{2} s_5 (\g_\mu \g^\nu)_a{}^b \partial_\nu \lambda_b + s_0(\g^\nu \g_\mu)_a{}^b \partial_\nu 
\lambda_b  + (\g^\nu)_{a}{}^{b} \partial_{[\mu} \psi_{\nu ] b}  \\[6pt]
{\rm D}_a A_\mu & \, =  i \frac{1}{2}(\g^5\g_\mu)_a{}^b (\chi_b + s_4 R_b) + i \frac{1}{2} s_7(\g^5 \g_\mu \g^\nu)_a{}^b \partial_\nu \lambda_b +  i \frac{1}{2} s_{0}(\g^5 \g^\nu \g_\mu)_a{}^b \partial_\nu \lambda_b  \cr
&\hspace{15 pt}+ \frac{1}{2} \epsilon_{\mu}^{~\nu\alpha\beta}(\g_\nu)_a^{~b}\partial_\alpha \psi_{\beta b} \\[6pt]
{\rm D}_a G_\mu & \, =  \epsilon_\mu^{~\nu\alpha\beta} (\g_\nu)_a{}^b \partial_\alpha 
\psi_{\beta b} - i s_{0}(\g^5[\g_\mu, \g^\nu])_a^{~b} \partial_\nu \lambda_b \\[6pt]
{\rm D}_a B_\mu & \, = \psi_{\mu a} + s_4 (\g_\mu)_a{}^b \lambda_b \\[6pt]
{\rm D}_a t_{\mu \nu} & \, = \frac{1}{4} ([\g_\mu,\, \g_\nu])_a{}^b (\chi_b + s_0R_b) - 2 s_4 (\g_{[\mu})_{|a|}^{~~b}\partial_{\nu]} \lambda_b  + \frac{1}{4} s_3 ([\g_\mu , \g_\nu]\g^\alpha)_a{}^b \partial_\alpha \lambda_b  \cr
&\hspace{15 pt}+ \partial_{[\mu} 
\psi_{\nu] a} + i \epsilon_{\mu\nu}{}^{\alpha\beta}(\g^5)_a{}^b \partial_\alpha \psi_{
\beta b} 
\end{align}
where as in the dWvH case, $R_a \equiv ([\g^\mu,\, \g^\nu])_a{}^b \partial_\mu  \psi_{\nu b}$ and $F_{\mu 
\nu} \equiv \partial_{[\mu} B_{\nu]}$  and
\begin{align}
	s_1 =& 3 s_0^2 - 1 ,~~~s_2 = 3 s_0^2 + 6 s_0 +1 ,~~~s_3 = 3 s_0^2 + 2 s_0 +1,~~~s_4 = s_0 +1 \cr
	s_5 =& 3 s_0^2 + 4 s_0 +1 ,~~~s_6 = 3 s_0^2 - 2 s_0 -3 ,~~~s_7 = 3 s_0^2 + 5 s_0 +1
\end{align}

\subsection{Anti-Commutators}
The Algebra for the OS multiplet is as follows:
\begin{align}
\{ {\rm D}_a,\, {\rm D}_b \}X \, = & 2 i (\g^{\mu})_{ab} \partial_{\mu} X ~~ \mbox{for} ~~ 
X \in \left\{ \lambda_c,\, \chi_c,\, P,\, V_\mu,\, G_\mu,\, t_{\mu \nu} \right\} \\[6pt]
\{ {\rm D}_a,\, {\rm D}_b \} A_\mu \, = & 2 i (\g^{\alpha})_{ab} \partial_{\alpha} A_{\mu} 
- \partial_\mu \xi_{ab} \\[6pt]
\{ {\rm D}_a,\, {\rm D}_b \} B_\mu \, = & 2 i (\g^{\alpha})_{ab} \partial_{\alpha} B_{\mu} 
- \partial_\mu \Lambda_{ab} \\[6pt]
\{ {\rm D}_a,\, {\rm D}_b \} \psi_{\mu c} \, = & 2 i (\g^\alpha)_{ab} \partial_\alpha \psi_{
\mu c} - \partial_\mu \epsilon_{abc}
\end{align}
with $\Lambda_{ab}$ and $\epsilon_{abc}$ as in Equation~(\ref{e:dWvHNonClosure}) with $c_0 \to s_0$ and the new gauge term 
\begin{align*}
	\xi_{ab} =& 2 i (\g^\nu)_{ab} \left( A_\nu - \frac{1}{2} G_\nu\right).
\end{align*}
The algebra closes on the field strengths
\begin{align}
	\{ {\rm D}_a,\, {\rm D}_b \} f_{\mu\nu} \, = & 2 i (\g^\alpha)_{ab} \partial_\alpha f_{\mu\nu}  \cr
	\{ {\rm D}_a,\, {\rm D}_b \} F_{\mu\nu} \, = & 2 i (\g^\alpha)_{ab} \partial_\alpha F_{\mu\nu}  \cr
	\{ {\rm D}_a,\, {\rm D}_b \} R_{c} \, = & 2 i (\g^\alpha)_{ab} \partial_\alpha R_{
c} ~~~
\end{align}
where $f_{\mu\nu} = \partial_\mu A_\nu - \partial_\mu A_\nu$.
\subsection{Lagrangian}

The Lagrangian that is invariant with respect to the OS transformation laws is
\begin{align}
\mathcal{L}  \, = & - P^2 + \frac{1}{2} V_\mu V^\mu + A_\mu 
G^\mu - \frac{1}{2} F_{\mu\nu}F^{\mu\nu} + \frac{1}{4} t_{\mu\nu}t^{\mu\nu} - \frac 12 \,  \psi_{\mu a} \epsilon^{\mu\nu\alpha\beta} (\g^5 \g_\nu)^{ab} 
\partial_\alpha \psi_{\beta b} +  i \lambda_b \chi^b.
\end{align}

As in the dWvH case, the OS Lagrangian is invariant with respect to the fermionic field redefinition, as in~Equation~(\ref{e:fermredef}) with $c_0 \to s_0$.
The OS Lagrangian is also invariant with respect to the same gauge transformations as the dWvH case, Equation~(\ref{e:dWvHgt}).

\newpage
\section{The Non-Minimal Supergravity Formulation}\label{s:nmSG}
The dynamical field content of \nmSG~ is that of the graviton $h_{\mu\nu}$, which is symmetric but \emph{not} traceless in our formulation, and the gravitino $\psi_{\mu a}$, which is likewise not traceless. The auxiliary field content for \nmSG~ consists of a scalar field $S$, pseudoscalar field $P$, two pseudovector fields $A_\mu$ and $W_\mu$, a vector field $V_\mu$, and two spinors $\lambda_a$ and $\beta_a$, the former being a leading order fermion in the superfield expansion. The transformation laws, algebra, and Lagrangian for \nmSG~ are given in the following~subsections.


\subsection{Transformation Laws}

We write the transformation laws in terms of a single free parameter $g_0$, which parameterizes a field redefinition of the fermionic fields as in Equation~(\ref{e:fermredef}) but with $c_0 \to g_0$ and $\chi_a \to \beta_a$ that leaves the Lagrangian invariant. The other parameter $n$ in the \nmSG~multiplet is a remnant from the superspace formulation of supergravity where $n=-1/3$ reduces the formulation to the first minimal, off-shell version of 4D, $\mathcal{N}=1$  supergravity discovered, sometimes referred to as old-minimal supergravity, and $n=0$ to the next, sometimes referred to as new-minimal supergravity~\cite{SFSG}.
\begin{subequations}\label{e:DNMEasy}
\begin{eqnarray}
D_a S &=& \frac{1 + 3 n}{4n} \beta_a - g_1 R_a + g_2 (\g^\nu)_a^{~b}\partial_\nu \lambda_b\\*
D_a P &=&  -i  (\g^5)_a{}^b (D_b S) \\
D_a A_\mu &=& i (\g^5 \g^\nu)_a{}^b \partial_{[\nu} \psi_{\mu]b} - 
\frac{1}{2} \epsilon_\mu{}^{\nu\alpha\beta} (\g_\nu)_a{}^b \partial_{
\alpha} \psi_{\beta b} + i g_3  (\g^5)_a{}^b \partial_\mu \lambda_b  \\
D_a h_{\mu\nu} &=& \frac 12 \,  (\g_{(\mu})_{a}{}^b \psi_{\nu)b}  - g_0 \eta_{\mu\nu} \lambda_a \\
D_a V_\mu &=& -\frac{1}{4} (\g_\mu)_a{}^b (\beta_b - g_0 R_b) - \frac{3}{4}g_0^2 (\g_\mu \g^\nu)_a{}^b \partial_\nu \lambda_b - \frac{n}{1+3n}  (\g^\nu 
\g_\mu)_a{}^b \partial_\nu \lambda_b \\
D_a W_\mu &=& i \frac{1}{6}  (\g^5 \g_\mu)_a{}^b (3 \beta_b - g_3 R_b)  + i \frac{1}{6} g_3^2 (\g^5 \g_\mu \g^\nu)_{a}^{~b}\partial_\nu \lambda_b  + i \frac{2}{3 (1 + 3n)} (\g^5 \g^\nu \g_\mu )_{a}^{~b}\partial_\nu \lambda_b   \\
D_a \psi_{\mu b} &=&  i 2 n g_1 (\g_\mu)_{ab} S + 2 n g_1 (\g^5 \g_\mu)_{ab} P + 
\frac{2}{3} (\g^5)_{ab} A_\mu + \frac{1}{6} (\g^5 [\g_\mu, \g^\nu])_{ab} A_\nu \cr
&&- i \frac{1}{2} [ \g^\alpha, \g^\beta]_{ab} \partial_\alpha h_{\beta\mu}  + \frac{1+3n}{4}  g_3 (\g^5\g^\nu\g_\mu)_{ab} W_\nu + i \frac{1 + 3n}{2n} g_0 (\g^\nu \g_\m)_{ab} V_\n  \\
D_a \lambda_b &=& -(1+3n)\left( i \frac{1}{2} C_{ab} S + \frac{1}{2} (\g^5)_{ab} P + \frac{3}{4} (\g^5 \g^\mu
)_{ab} W_\mu + i \frac{1}{2n} (\g^\mu)_{ab} V_\mu  \right) \\
D_a \beta_b &=& i  2 n g_2  (\g^\mu)_{ab} \partial_\mu S + 2 n g_2 (\g^5 \g^\mu)_{ab} 
\partial_\mu P + (\g^5\g^\mu\g^\nu)_{ab} \partial_\mu W_\nu + \frac{1}{4} g_3^2(1+3n)  (\g^5\g^\nu\g^\mu)_{ab} \partial_\mu W_\nu  \cr
&&+ 2 i (\g^\mu \g^\nu)_{ab} \partial_\mu V_\nu + i \frac{3}{2} g_0^2 \frac{1 + 3n}{n} (\g^\n \g^\m)_{ab} \partial_\m V_\n  - \frac{2}{3}g_3 (\g^5)_{ab} \partial_\mu A^\mu \cr
&& -2 i g_0 C_{ab} (\square h - \partial_\mu \partial_\nu h^{\mu\nu})
\end{eqnarray}
\end{subequations}
where
\begin{align}
	g_1 =&  \frac{g_0 + (2 + 3 g_0)n}{4 n},~~~g_2 = 3 g_0^2 \frac{1 + 3 n}{4n} + 3 g_0 + 1,~~~g_3 = 3 g_0 + 2,~~~g_4 = 9 g_0^2 + 12 g_0 + 2
\end{align}

As before, the field strength of the gravitino is given by $R_a \equiv ([\g^\mu,\, \g^\nu])_a{}^b \partial_\mu  \psi_{\nu b}$.

\subsection{Anti-Commutators}
The algebra closes on the auxiliary fields $X = (S,P, A_\mu, V_\mu, 
W_\mu, \lambda_a, \beta_a)$ as
\begin{align}
\{ {\rm D}_a , {\rm D}_b \} X = 2 i (\g^\mu)_{ab} \partial_\mu X.
\end{align}
The algebras for the physical fields $\psi_{\mu a}$ and $h_{\mu\nu}$ are
\begin{align}\label{e:nmSGnonclosure}
\{ {\rm D}_a , {\rm D}_b \} h_{\mu\nu} = & 2 i (\g^\alpha)_{ab} \partial_\alpha 
h_{\mu\nu} - \partial_{(\mu} \zeta_{\nu) ab} \cr
\{ {\rm D}_a , {\rm D}_b \} \psi_{\mu c} = & 2 i (\g^\alpha)_{ab} \partial_\alpha  
\psi_{\mu c} - \partial_\mu \varepsilon_{abc}.
\end{align}
The gauge terms $\zeta_{\nu ab}$ and $\varepsilon_{abc}$  are
\begin{align}
\zeta_{\nu ab} =~& i (\g^\alpha)_{ab} h_{\nu \alpha} \\
\varepsilon_{abc} =~&  i \frac{1}{8} (10 (\g^\a)_{ab} \delta_c{}^d - [ \g^\a , \g^\b]_{ab} (\g_\b)_c{}^d + (\g_\b)_{ab} ([\g^\b , \g^\a])_c{}^d - (\g^5 [\g^\a , \g^\b])_{ab} (\g^5 \g_\b)_c{}^d )\psi_{\a d} \cr
& - i \frac{1}{8}( (16 g_0 + 8) (\g^\a)_{ab} (\g_\a)_c{}^d + (g_0 + 1) ([\g^\a , \g^\b])_{ab} ([\g_\a , \g_\b])_c{}^d ) \l_d
\end{align}

The algebra closes on the field strengths
\begin{align}
	\{ {\rm D}_a, {\rm D}_b \} \mathcal{R}_{\alpha\mu\beta\nu} = &  2 i (\gamma^\rho)_{ab} \partial_\rho \mathcal{R}_{\alpha\mu\beta\nu}, \cr
		\{ {\rm D}_a, {\rm D}_b \} R_{c} = &  2 i (\gamma^\alpha)_{ab} \partial_\alpha R_{c}.
\end{align}
where the weak field Riemann tensor $\mathcal{R}_{\alpha\beta\mu\nu}$ is 
\begin{align}
\mathcal{R}_{\alpha\mu\beta\nu} =& \frac{1}{2}(\partial_{\mu}\partial_\nu h_{\alpha\beta} - \partial_{\mu}\partial_\beta h_{\alpha\nu} + \partial_{\alpha}\partial_\beta h_{\mu\nu} - \partial_{\alpha}\partial_\nu h_{\mu\beta}).
\end{align}
\subsection{Lagrangian}

The Lagrangian for \nmSG is
\begin{align}\label{e:LNMSimple}
\mathcal{L}_{\text{\nmSG}} =& -\, \frac 12 \,  \partial_\alpha h_{\mu\nu} \partial^\alpha 
h^{\mu\nu} + \frac 12 \partial_\alpha h \partial^\alpha h -  \partial^\alpha h \partial^\beta 
h_{\alpha\beta} +  \partial^\mu h_{\mu\nu} \partial_\alpha h^{\alpha\nu} + n 
S^2  +   n P^2 +\frac{ 1}{3}  A_\mu A^\mu \cr
&- \frac{1+3n}
{n}  V_\mu V^\mu  - \frac{3}{4}(3 n+1)  W_\mu W^\mu - \frac 12 \,  \psi_{\mu a} \epsilon^{\mu\nu\alpha
\beta} (\g^5 \g_\nu)^{ab} \partial_\alpha \psi_{\beta b} + i \lambda_a 
\beta^a.
\end{align}
As in the dWvH and OS cases, the \nmSG~Lagrangian is invariant  with respect to the field redefinitions as in Equation~(\ref{e:fermredef}) but with $c_0 \to g_0$ and $\chi_a \to \beta_a$. The \nmSG~Lagrangian is also invariant with respect to the following gauge transformations that are indicated by Equation~(\ref{e:nmSGnonclosure})
\begin{subequations}
\begin{align}\label{e:nmSGgt}
	\delta h_{\mu\nu} = & \partial_{(\mu} \zeta_{\nu)} \\
	\delta \psi_{\mu a} = & \partial_\mu \varepsilon_a.
\end{align}
\end{subequations}
The fermionic part of the \nmSG~Lagrangian  is identical to those of OS and dWvH under the identification $\beta_a = \chi_a$.

\newpage


\section{Adinkranization of the \boldmath{$20 \times 20$} Multiplets}\label{s:adinkranization}
Here, we summarize the adinkranization process. More details can be found in 
the appendices. Considering the fields in the dWvH, OS, and \nmSG ~multiplets 
to be only time dependent, we gauge fix to temporal gauge
\begin{align}
	\psi_{0 a} =& B_0 =  0,~~~\text{dWvH}\\
	\psi_{0 a} =& B_0 = A_0  = G_0 = 0,~~~\text{OS}\\
	\psi_{0 a} =& A_0 = h_{0\mu} = 0,~~~\text{\nmSG}
\end{align} 


Expanding on the discussion in Section~\ref{s:GadgetReview} regarding the smaller CM, TM, and VM multiplets, we define a convention for nodal field definitions that is consistent with the CM, TM, and VM that can be applied to the larger $20 \times 20$ multiplets. First, dynamical fields appear to the left of auxiliary fields. For auxiliary fermions, those of lower mass dimension appear to the left of those of higher mass dimension. For bosonic fields, they are listed in the nodes left to right in the following order: scalars, pseudoscalars, vectors, pseudovectors, tensors, and pseudotensors. Gauge fields appear to the right of non-gauge fields of the same rank. In the case of multiple pseudoscalars for instance, the pseudoscalar that comes in a pair with a scalar (that form a complex scalar as in the fields$K$ and $L$ of the dWvH multiplet for instance) appears before non-paired pseudoscalars. Fields with components are listed left to right in numerical order if there is a single component. Fields with more complicated index structure, such as the graviton, gravitino, and antisymmetric tensors, are listed in the orders shown in the specific examples below.

For the dWvH formulation of the ($\frac32$,1) supermultiplet, we order the bosons according to
\be
\Phi_i ~=~ 
\left(
B_1 , \,
B_2 , \,
B_3 , \,
\int K ~dt , \,
\int L ~dt  , \,
\int P ~dt  , \,
\int V_{\mu}~dt   , \,
\int U_{\mu}~dt   , \,
\int t_{\mu \, \nu}~dt  
\right),
\ee
for the OS formulation we order the bosons according to
\be
\Phi_i ~=~ \left(
B_1 , \,
B_2 , \,
B_3 , \,
\int P ~dt , \,
\int V_{\mu}~dt  , \,
\int A_{1}~dt  , \,
\int A_{2}~dt  , \,
\int A_{3}~dt  , \,
\int t_{\mu \, \nu}~dt  , \,
E_{1 \, 2} , \,
E_{2 \, 3} , \,
E_{3 \, 1} \right) 
\ee
where the ordering for $t_{\mu\nu}$ is as follows for both the dWvH and OS multiplets:
\begin{align}
	\{t_{01}, t_{02}, t_{03}, t_{12}, t_{23}, t_{31}\}
\end{align}
Note that
\begin{align}
	E_{12} = \int G^3 dt = \int G_3 dt,~~~E_{23} = \int G^1 dt = \int G_1 dt ,~~~E_{31} = \int G^2 dt = \int G_2 dt
\end{align}
Finally, for the non-minimal SG bosons,
\be
\Phi_i ~=~ \left(
h_{11} , \,
h_{12} , \,
h_{13} , \,
h_{22} , \,
h_{23} , \,
h_{33} , \,
\int S ~dt , \,
\int P ~dt , \,
\int V_{\mu}~dt  , \,
\int W_{\mu}~dt  , \,
\int A_{\mu}~dt 
\right). 
\ee

Next, for both dWvH and OS formulations fermions, we choose
\be
i\Psi_i ~=~ 
\left(
\psi_1 {}_a , \,
\psi_2 {}_a , \,
\psi_3 {}_a , \,
\int\lambda_a ~dt , \,
\int\chi_a ~dt 
\right),
\ee
while, for fermions of the non-minimal SG supermultiplet fermions, we use
\be
i\Psi_i ~=~ 
\left(
\psi_1 {}_a , \,
\psi_2 {}_a , \,
\psi_3 {}_a , \,
\int\lambda_a ~dt , \,
\int\beta_a ~dt 
\right).
\ee

With these definitions, the transformation laws for each multiplet can be succinctly written as 
 \begin{align}
{\rm D}_{\rm I} \Phi = i {\rm \textbf{L}}_{\rm I} \Psi,~~~{\rm D}_{\rm I} \Psi = {
\rm \textbf{R}}_{\rm I} \dot{\Phi}.
\end{align}

As it is not terribly instructive to display all $\brL_\rI$ and $\brR_\rI$ matrices for all of these multiplets we have published them along with all of the adinkra data described below for these three multiplets in three \emph{Mathematica} data files \emph{dWvH.m}, \emph{OS.m}, and \emph{nmSG.m} at the \href{https://hepthools.github.io/Data/}{Data repository} on GitHub. A master file\emph{Compare20x20Reps.nb} is located at the same repository which demonstrates how to display the data and perform the various calculations summarized in the remainder of the paper. The tutorial file  \emph{Compare20x20Reps.nb} utilizes the \emph{Mathematica} package \href{https://hepthools.github.io/Adinkra/}{\emph{Adinkra.m}}, which is available at a different GitHub Repository.  A   general tutorial \emph{AdinkraTutorial.nb} that demonstrates the various features of the \emph{Adinkra.m} package is also located at the \href{https://hepthools.github.io/Adinkra/}{\emph{Adinkra.m} repository}.

In Appendix~\ref{a:LRmatrices}, we display the explicit  $\brL_\rI$ and $\brR_\rJ$ matrices for the $c_0 = 0$ representation of the dWvH multiplet. 
For all three multiplets, the $\brL_\rI$ and~$\brR_\rJ$ matrices satisfy the $GR(d,N)$ algebra, the~algebra of general, real matrices of size $d\times d$ that encode $N$ supersymmetries~\cite{G-1}: 
\begin{align}
	{\rm \textbf{L}}_{\rm I} {\rm \textbf{R}}_{\rm J} +{\rm \textbf{L}}_{\rm J} {\rm \textbf{R}}_{\rm I} = & ~2 \delta_{\rm IJ} {\rm \textbf{I}}_d \\
	{\rm \textbf{R}}_{\rm I} {\rm \textbf{L}}_{\rm J} +{\rm \textbf{R}}_{\rm J} {\rm \textbf{L}}_{\rm I} = & ~2 \delta_{\rm IJ} {\rm \textbf{I}}_d
\end{align}
As $d=20$ and $N=4$ for the dWvH, OS, and \nmSG~multiplets, their $\brL_\rI$ and$\brR_\rJ$ matrices satisfy more specifically the $GR(20,4)$ algebra.

Recall, the parameter $\chi_0$ is defined through the relationship
\begin{align}
Tr(\brL_{\rI}\brR_{\rJ}\brL_{\rK}\brR_{\rL}) = 4\left[ (n_c + n_t) (\delta_{IJ}\delta_{KL
} - \delta_{IK}\delta_{JL} + \delta_{IL}\delta_{JK}) + \chi_0 \epsilon_{IJKL} \right]
\end{align}
The parameters $n_c$ and $n_t$ are referred to as the isomer parameters. They 
encode the number $n_c$ cis-isomer adinkras and the number $n_t$ trans-isomer 
adinkras into which a multiplet can be decomposed. The parameter $\chi_0 = n_c 
- n_t$.  
For the dWvH, OS, and \nmSG multiplets, we find 
\be\label{e:chi0}
\eqalign{
	\chi_0 = 3,~~~ n_c &= 4,~~~n_t = 1~~~~~\text{dWvH} \cr
	\chi_0 = 1,~~~n_c &= 3,~~~n_t = 2~~~~~\text{OS} \cr
	\chi_0 = -3,~~~n_c &= 1,~~~n_t = 4~~~~~\text{\nmSG}}
\ee

\subsection{Holoraumy and \texorpdfstring{$so(4) = su(2) \times su(2) $}{so(4)=su(2)xsu(2)}} 
Recall 
the matrix representations for fermionic and bosonic holoraumy are defined as 
\begin{align}
\brV_{\rI\rJ} =& - \frac{i}{2} \textbf{L}_{[I} \textbf{R}_{J]},~~~
\brtV_{\rI\rJ} = - \frac{i}{2} \textbf{R}_{[I} \textbf{L}_{J]}.
\end{align}
For any set of matrices  $\brL_\rI$ and $\brR_\rJ$ that satisfy the $GR(d,N)$ algebra, Equation~(\ref{e:GRdN}), setting either $\brV_{\rI\rJ} = 2 t_{\rI\rJ}$ or $\brtV_{\rI\rJ} = 2 t_{\rI\rJ}$ will satisfy the so(N) algebra
\begin{align}\label{e:soN}
[t_{IJ} , t_{KL}] ~=~ i (\delta_{I[L} t_{K]J}-\delta_{J[L} t_{K]I}).
\end{align}
A proof is given in Appendix~\ref{a:soNProof}. 

For the special case of $so(4)$, we define 
\begin{align}\label{e:Vpm}
\brV^{\pm}_{\rm IJ} ~\equiv&~ \frac{1}{2} \left(\brV_{\rm IJ} \pm \, \fracm 12 \, \epsilon_{\rm IJKL} \brV_{\rm KL} 
\right)~~~ \\
\label{e:Vtildepm}
\brtV^{\pm}_{\rm IJ} ~\equiv&~ \frac{1}{2} \left(\brtV_{\rm IJ} \pm \, \fracm 12 \,  \epsilon_{\rm IJKL} \brtV_{\rm KL} \right)
\end{align}
where Einstein summation convention is assumed on the repeated indices $K$ 
and $L$. It is straightforward to show that both $t_{\rm IJ} = 1/2\brV^{\pm}_{IJ}$ and $t_{\rm IJ} = 1/2\brtV^{\pm}_{IJ}$ satisfy the $so(4)$ algebra, Equation~(\ref{e:soN}). At the same time, $\brV^\pm_{IJ}$ and  $\brtV^\pm_{IJ}$   only have 
three independent elements each. We display the independent elements of $\brV^\pm_{\rm IJ}$ below; those of $\brtV^\pm_{\rm IJ}$ satisfy similar relations:
\begin{align*}
\brV^+_{12} ~=~ \brV^+_{34},~~~\brV^+_{13} ~=~ \brV^+_{42},~~~
\brV^+_{23} ~=~ \brV^+_{14}, \cr
\brV^-_{12} ~=~ \brV^-_{43},~~~\brV^-_{13} ~=~ \brV^-_{24},~~~
\brV^-_{23} ~=~ \brV^-_{41},
\end{align*} 

Furthermore, all $\brV^+_{IJ}$ commute with all $\brV^-_{IJ}$. 
In this way, $\brV^\pm_{IJ}$ are actually two separate, commuting representations of 
$su(2)$:
\begin{align*}
[\brV^+_{12},\brV^+_{13}] =~& 2 i \brV^+_{23},~~~~[\brV^+_{13},\brV^+_{23}] = 2 i \brV^+_{12}
,~~~~[\brV^+_{23},\brV^+_{12}] = 2 i \brV^+_{13},\cr
[\brV^-_{12},\brV^-_{13}] =~& 2 i \brV^-_{23},~~~~[\brV^-_{13},\brV^-_{23}] = 2 i \brV^-_{12}
,~~~~[\brV^-_{23},\brV^-_{12}] = 2 i \brV^-_{13},\cr
[\brV^+_{IJ},\brV^-_{KL}] =~& 0.
\end{align*}
Similar relationships are satisfied by $\brtV^{\pm}_{\rm IJ}$.

\subsection{\texorpdfstring{$\brtV_{\rm IJ}$}{VIJ}, Eigenvalues, and Gadgets for the dWvH, OS, and \texorpdfstring{\nmSG}{nmSG} Multiplets}\label{s:Gadgets}

The explicit matrix forms of $\brV_{\rm IJ}$ and $\brtV_{\rm IJ}$ are too large to display and be instructive in this paper. We have published them open-source in the files \emph{dWvH.m}, \emph{OS.m}, and \emph{nmSG.m} at the previously mentioned GitHub \href{https://hepthools.github.io/Data/}{data repository}. As an example, in Appendix~\ref{a:tildeVmatrices} we show the explicit form for the $\brtV_{\rm IJ}$ for the $c_0 = 0$ representation of the dWvH multiplet. Unlike the fundamental $CM$,  $TM$, and $VM$ representations \cite{KIAS1,KIAS2,H4,H3,G&G1,G&G2}, the $\brV_{\rm IJ}$ and $\brtV_{\rm IJ}$ for the dWvH, OS, and \nmSG~ representations are all true $so(4)$ representations composed of six linearly independent elements:
\begin{align*}
\brV_{12},&~~~\brV_{13},~~~\brV_{14},~~~\brV_{23},~~~
\brV_{24},~~~\brV_{34},\cr
\brtV_{12},&~~~\brtV_{13},~~~\brtV_{14},~~~
\brtV_{23},~~~\brtV_{24},~~~\brtV_{34}.
\end{align*}

In contrast, the $\brV_{\rm IJ}$ and $\brtV_{\rm IJ}$ for the $CM$, $VM$, and $TM$ each form a single, non-trivial $su(2)$ representation, with only three linearly independent 
algebra elements \cite{KIAS1,KIAS2,H4,H3,G&G1,G&G2}.  That is either the $\brtV^+_{\rm IJ}$ or the $\brtV^-_{\rm IJ}$ vanish and either the either the $\brV^+_{\rm IJ}$ or the $\brV^-_{\rm IJ}$ vanish for the $CM$, $VM$, and $TM$. This is not the case for the dWvH, OS, and \nmSG~ representations: the $\brtV^\pm_{\rm IJ}$ 
for these are all nontrivial. We see then for the 
dWvH, OS, and \nmSG~ representations, the $\brV_{\rm IJ}$ and $\brtV_{\rm IJ}$ all 
form true $so(4)$ representations, each which separate into two commuting $su(2)$ 
representations, $\brV^\pm_{\rm IJ}$ and $\brtV^\pm_{\rm IJ}$, respectively, as shown 
in the previous section. The eigenvalues for $\brV_{\rm IJ}$ and $\brtV_{\rm IJ}$ for the dWvH, OS, and \nmSG ~multiplets 
are all~$\pm 1$. 

{All of the dWvH, OS, and \nmSG~multiplets have gadgets, Equation~(\ref{e:Gadget}), that are normalized to 
\\$5 = n_c + n_t$:}
\begin{align}
\mathcal{G}(\text{dWvH},\text{dWvH}) ~= & ~5,~~~
\mathcal{G}(\text{OS},\text{OS}) ~= ~5 ,~~~
\mathcal{G}(\text{\nmSG},\text{\nmSG}) = 5.
\end{align}
The gadgets between the three different representations depend on the diagonal Lagrangian parameters $c_0$, $s_0$, $g_0$ as well as the superspace supergravity parameter $n$. While presenting the results below, we comment on the interesting cases where gadgets between the different representations are zero or five. As described in Section~\ref{s:GadgetReview}, where the gadget is described as the vector analogy of a dot product, a gadget of zero means the multiplets are gadget-orthogonal, which is analogous to two vectors being orthogonal. A gadget value of $5 = n_c + n_t$ is analogous to two vectors being parallel.

{
First, we define the self-gadget of a representation as the gadget between the same representation with two different values of its Lagrangian parameter: one unprimed, the other primed. We then have the following three sets of parameters to consider, one set for each of the $20 \times 20$ representations: ($c_0,~c_0'$), ($s_0,~s_0'$), and ($g_0,~g_0'$). We find the following self-gadget values:
\begin{align}
	\mathcal{G}(\text{dWvH},\text{dWvH'}) =& ~5 + 9/2 (c_0 - c_0')^4 + 2 (c_0 - c_0')^2 \\
	\mathcal{G}(\text{OS},\text{OS'}) =& ~5 + \frac{21}{8} (s_0 - s_0')^4 - 3 (s_0 - s_0')^2 \\
	\mathcal{G}(\text{\nmSG},\text{\nmSG'}) =& ~5 + (g_0 - g_0')^2 \frac{(3 n + 1)^2 }{2n^2} \left(\frac{9}{16} (9 n^2 + 2 n + 1)  (g_0 - g_0')^2 - n   \right)
\end{align}

This demonstrates interestingly that five is the \emph{minimum} value that the dWvH self-gadget can take. The minimum self-gadget value for the OS multiplet is precisely $29/7 \approx 4.14286$ and the minimum value for the \nmSG~self-gadget is $14/3 \approx 4.66667$. The OS self-gadget equals five for three separate relationships between $s_0$ and $s_0'$. The \nmSG~self-gadget equals five for the precise value of $n=-1/3$, two solutions of $n$ that depend on $g_0$ and $g_0'$, and of course the case $g_0 = g_0'$. The self-gadgets are summarized in Table~\ref{t:SelfGadgets} where to more succinctly write the \nmSG~results, we define the function
\begin{align}\label{e:rpm}
	r_{\pm}(x) = (8 - 9 x^2 \pm 2 \sqrt{2} \sqrt{ 8 - 18 x^2 - 81 x^4} )/(81 x^2)
\end{align}
}
\noindent
{
It is worth noting that the minimum case $ n = -1/3$ for \nmSG~corresponds to its reduction to old-minimal supergravity~\cite{SFSG}, as described in Section~\ref{s:nmSG}.
}

\renewcommand{\arraystretch}{1.5}
\begin{table}[!htbp]
\caption{Self gadgets between the different multiplets. The function $r_{\pm}(x)$ is defined in Equation~(\ref{e:rpm}) .}
\centering
\begin{tabular}{cccc}
\toprule
\textbf{Multiplet} & \textbf{Minimum} & \textbf{When Minimum} & \textbf{When Equals Five} \\
\midrule
dWvH & 5  & $c_0 = c_0'$ & $c_0 = c_0'$\\
OS & 29/7 & $s_0 - s_0' \approx 0.755929$ & $s_0 - s_0' = 0,~ \pm 2 \sqrt{2/7} $ \\
\nmSG & 14/3 &  $n = g_0 - g_0' = 1/3$ & $g_0 = g_0'$ or $ n = -1/3,~ r_{\pm}(g_0 - g_0')$  \\
\bottomrule 
\end{tabular}
\label{t:SelfGadgets}
\end{table}

The gadgets between the dWvH, OS, and nmSG multiplets are as follows: 
\begin{align}\label{e:dWvHOSGadget}
	\mathcal{G} (\text{dWvH},\text{OS}) = &(-c_0+s_0-1) \left(-3 c_0^3+\left(9 c_0^2+1\right) s_0-9
   c_0 s_0^2-c_0+3 s_0^3+1\right)+5 \cr
   =& ~3 (c_{0} - s_{0})^{4} + 3 (c_{0} - s_{0})^{3} + (c_{0} - s_{0})^{2} + 4\\
   \label{dWvHnmSGGadget}
   	\mathcal{G} (\text{dWvH},\text{\nmSG}) =&  - \frac{1}{4n} \Big\{   3 (3n+1)^{2} (c_{0} + g_{0})^{4}   + 6 (3n+1)(7n+1) (c_{0} + g_{0})^{3}  \cr
    & \qquad  + 2 (9n+1)(13n+3) (c_{0} + g_{0})^{2}   + 2 (97n^{2} + 22n + 1) (c_{0} + g_{0})  \Big\}  \cr
    &  - \frac{1}{12n} (181n^{2} + 26n + 1) \\
    \label{OSnmSGGadget}
    \mathcal{G} (\text{OS},\text{\nmSG}) =&  \frac{1}{16n} \Big\{   -9 (n+1)(3n+1) (s_{0} + g_{0})^{4}   + 12 (n-1)(3n+1) (s_{0} + g_{0})^{3}  \cr
    & \qquad  + 2 (153n^{2} + 52n - 5) (s_{0} + g_{0})^{2}  + 4 (91n^{2} + 30n - 1) (s_{0} + g_{0})  \Big\}  \cr
    &  + \frac{1}{48n(3n+1)} (1077n^{3} + 531n^{2} + 59n - 3)
\end{align}
Upon closer inspection of these gadgets, we find some interesting facts as to holographic possibilities. For instance, an obvious solution for which dWvH and OS are parallel, i.e., have a gadget value equal to five, is
\begin{equation}\label{e:dWvHOSparallel}
	\mathcal{G} (\text{dWvH},\text{OS}) = 5~~~\text{for}~~~s_{0} = c_{0} + 1
\end{equation}

The form of the gadget between dWvH and OS on the second line of Equation~(\ref{e:dWvHOSGadget}), however, indicates perhaps a more natural choice might be
\begin{align}
	\mathcal{G} (\text{dWvH},\text{OS}) = 4~~~\text{for}~~~s_{0} = c_{0}
\end{align}

Solutions exist to make dWvH parallel to \nmSG, and OS parallel to \nmSG, but these solutions are complication conditional solutions on $n$ so we have published these calculations in the file \emph{Compare20x20Reps.nb}  at the previously mentioned GitHub {data repository}. Two obvious cases to investigate are $c_0 = -g_0$ and $s_0 = - g_0$, for which we find
\begin{align}
	\mathcal{G} (\text{dWvH},\text{\nmSG}) =& ~5~~~\text{for} ~~c_0 = -g_0~~\text{and}~~n= -0.463211 ~~\text{or}~~n = -0.0119273 \\
	\mathcal{G} (\text{OS},\text{\nmSG}) =& ~5~\text{for} ~s_0 = -g_0~\text{and}~ n=-0.321012~,~~n=-0.0169016~,~~ \text{or}~ n=0.513401
\end{align}

As to orthogonality (gadget value of zero), inspection of Equation~(\ref{e:dWvHOSGadget}) reveals that there are no real solutions for $c_0$ and $s_0$ that make the dWvH and OS multiplets orthogonal
\begin{equation}
	\mathcal{G} (\text{dWvH},\text{OS}) \ne 0~~~\text{for} ~c_0, s_0 \in \text{Reals.}
\end{equation}
We do have, however, that 
\begin{equation}
	\mathcal{G} (\text{dWvH},\text{\nmSG}) = 0~~~\text{for}~~~
	\begin{array}{c}
		g_{0} = - c_{0} + \frac{1}{3} ,~~~  n = - \frac{1}{9}  \\
		\text{or} \\
		g_{0} = - c_{0} - 1 ,~~~  n = - 1  
	\end{array}
\end{equation}

On the other hand, the OS and \nmSG~multiplets can be made to be orthogonal for various \emph{ranges} on $n$. As these solutions for OS-\nmSG~orthogonality are rather complicated and thus not terribly instructive in their entirety, we have published the results in the file \emph{Compare20x20Reps.nb}  at the previously mentioned GitHub {data repository}. An interesting case is the following where both the dWvH and OS multiplets each are simultaneously orthogonal to the \nmSG~ multiplet (but not each other):
\begin{equation}
    \mathcal{G} (\text{dWvH},\text{\nmSG}) =   \mathcal{G} (\text{OS},\text{\nmSG}) = 0~~~\text{for}~~~ g_0 = -s_0 - \frac{2}{3},~~~c_0 = s_0 +1,~~~\text{and}~~~ n = -\frac{1}{9}
\end{equation}
This leaves the obvious cases $c_0 = -g_0$ and $s_0=-g_0$ to investigate as to orthogonality. In~these cases, there is no real solution for dWvH-\nmSG~orthogonality and only one real solutions for OS-\nmSG~orthogonality:
\begin{align}
	\mathcal{G} (\text{dWvH},\text{\nmSG}) \ne & ~0 ~~~\text{for}~~~c_0 = - g_0~~~\text{and}~~~ n \in \text{Reals} \\
	\mathcal{G} (\text{OS},\text{\nmSG}) = & ~0 ~~~\text{for}~~~s_0 = - g_0~~~\text{and}~~~ n=0.0373449 
\end{align}

Finally, we summarize the dWvH-\nmSG~gadgets and OS-\nmSG~gadgets in the physically interesting cases of $n=-1$, $n=-1/3$, and $n=0$. In these cases, \nmSG~is known to reduce to a representation that is part of a tower of higher spin that extends to $\mathcal{N}=2$ SUSY~\cite{SFSG,Gates:1996xs,GK1,GK2,GK3}, old-minimal supergravity~\cite{SFSG}, and new-minimal supergravity~\cite{Buchbinder:2002gh}, respectively. Both gadgets $\mathcal{G} (\text{dWvH},\text{\nmSG})$ and $\mathcal{G} (\text{OS},\text{\nmSG})$ diverge for $n=0$ and  $\mathcal{G} (\text{OS},\text{\nmSG})$ diverges for $n=-1/3$, as shown in Figure~\ref{f:Plots}.
\begin{figure}[!htbp]
\centering
	\includegraphics[width = 0.5\textwidth]{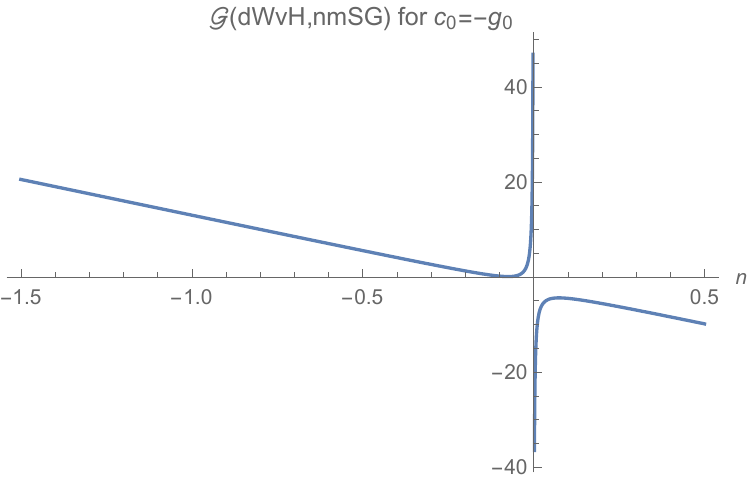}
	\quad
	\includegraphics[width = 0.4\textwidth]{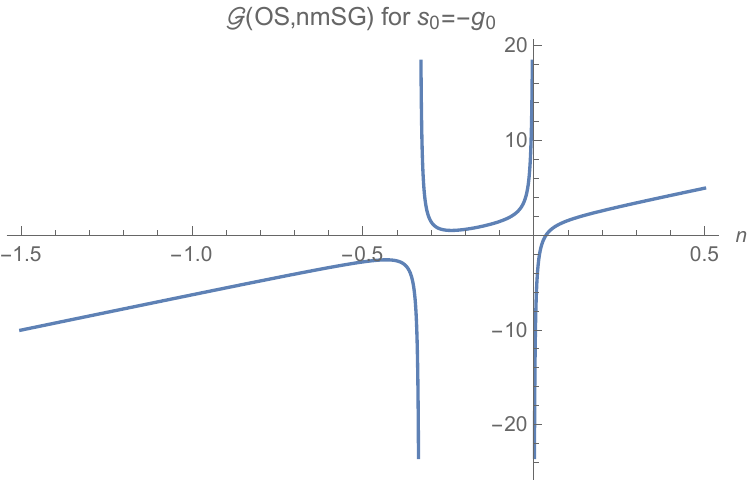}
	\caption{The gadgets $\mathcal{G} (\text{dWvH},\text{\nmSG})$ and $\mathcal{G} (\text{OS},\text{\nmSG})$ with $c_0 = -g_0$ and $s_0 = -g_0$.}
	\label{f:Plots}
\end{figure}

 Finite values of the gadget $\mathcal{G} (\text{dWvH},\text{\nmSG})$ exist for both $n= -1$ and $n=-1/3$ and a finite value for the gadget  $\mathcal{G} (\text{OS},\text{\nmSG})$ exists for $n=-1$.
\begin{align}
		\mathcal{G} (\text{dWvH},\text{\nmSG}) =&~ 3 \left(c_0+g_0\right){}^4+18 \left(c_0+g_0\right){}^3+40 \left(c_0+g_0\right){}^2+38
   \left(c_0+g_0\right)+13~\text{for}~n= -1 \\
   \mathcal{G} (\text{dWvH},\text{\nmSG}) =&~4 \left(c_0+g_0\right){}^2+\frac{20}{3} \left(c_0+g_0\right)+\frac{28}{9}~~~\text{for}~~~n= -1/3  \\
   \mathcal{G} (\text{OS},\text{\nmSG}) =& ~-3 \left(g_0+s_0\right){}^3-12 \left(g_0+s_0\right){}^2-15
   \left(g_0+s_0\right)-\frac{19}{3}~~~\text{for}~~~n=-1 
\end{align}

\noindent As these results along with Figure~\ref{f:Plots} indicate, the gadget values between these multiplets can be greater than the normalization of five. This is likely   from the non-adinkraic nature of the representations.

\newpage
\section{Conclusions}
\label{conclusions}
 In this paper, we investigated  three different 4D, $\mathcal{N}=1$ SUSY multiplets with \mbox{20 boson $\times$  20 fermion} degrees of freedom.  Specifically, we investigated  two matter gravitino multiplets, dWvH and OS, and non-minimal supergravity (\nmSG), each in a one-parameter family of component transformation laws that encode  an auxiliary fermion field redefinition symmetry of the diagonal Lagrangian. We furthered  research into SUSY genomics and holography by researching the dimensional reduction, $\chi_0$ values, and adinkra-level fermionic holoraumy of three $20 \times 20$ multiplets. All three have distinct $\chi_0$ values.  Gadgets calculated between the different multiplets indicate some interesting  possible connections to holography. The results in Section~\ref{s:Gadgets} demonstrate an elegant choice, Equation~(\ref{e:dWvHOSparallel}), for considering the OS and dWvH multiplets to be parallel in terms of the gadget, that is, have a gadget value of five.  Setting either the dWvH and \nmSG~multiplets parallel or the OS and \nmSG~multiplets parallel requires specific values of the supergravity parameter $n$ to be selected. On the other hand, \emph{no real solutions exist} that set the dWvH and OS multiplets orthogonal; however, at least one elegant solution exists that simultaneously sets both the dWvH and OS multiplets to be orthogonal to the \nmSG~multiplet. These results point to the possibility that holoraumy and the  gadget at the adinkra level indicate that the dWvH and OS multiplets are similar in some way, which we know in higher dimensions to be the case as they encode the same dynamical spins of $(3/2,1)$.
 
 Furthermore, fermionic holoraumy and the gadget seem to point to an adinkra-level distinction between the dWvH and \nmSG~multiplets and the OS and \nmSG~multiplets. We know of course that a key difference in 4D, $\mathcal{N}=1$ is that the dynamical fields of the \nmSG~ multiplet have spins $(2,3/2)$ rather than $(3/2,1)$ of the matter gravitino multiplets. We pointed  out some features of the gadgets for values of the supergravity parameter $n=-1$, $n=-1/3$ and $n=0$ which correspond to cases where \nmSG~ becomes part of a tower of higher spin that extends to $\mathcal{N}=2$ SUSY~\cite{SFSG,Gates:1996xs,GK1,GK2,GK3}, reduces to old-minimal supergravity~\cite{SFSG}, and reduces to new-minimal supergravity~\cite{Buchbinder:2002gh}, respectively. 
 
   A precise relationship between fermionic holoraumy and the gadget and spin of the higher dimensional system is still unknown. We look to uncover such precise spin--holography relationships not only through more research of  these $20 \times 20$ multiplets, but also into $12 \times 12$ multiplets of 4D, $\mathcal{N}=1$ supersymmetry, as well as higher spin mutliplets as in~\cite{Gates:1996xs,GK1,GK2,GK3}.  There are four $12 \times 12$ representations of 4D, $\mathcal{N}=1$ off-shell supersymmetry, and of these only one (CLS) has the fermionic auxiliary field redefinition symmetry similar to that presented in this work. This analysis is already being done and we hope to complete it soon. In addition, it would be interesting to see what other gadgets, such as those described in~\cite{Gates:2018pxb,Friend:2018ree}, encode for the $12 \times 12$, $20 \times 20$, and higher spin multiplets. 
In future works, with the $12 \times 12$ and higher spin multiplets, we look for more data to use with the $20 \times 20$ gadget data presented in this work to fix a canonical nodal field definition convention that remains consistent among all multiplets and perhaps to fix the diagonal Lagrangian parameters that will continue to be present in higher spin multiplets.


 \vspace{.05in}
 \begin{center}
{{\it ``The most effective way to do it, is to do it.'' \\ ${~}$ 
\\ ${~}$ }\hspace{300pt}\,\,-\,\, Amelia Earhart}
 \end{center}


  \noindent
{\bf Acknowledgments}\\[.1in] \indent
 This work was partially supported by the National Science Foundation grant PHY-1315155. This
research was also supported in part by the University of Maryland Center for String \& Particle Theory (CSPT). K.S.   thanks  Northwest Missouri State University for computing equipment and travel funds, and Dartmouth College and the E.E. Just Program for hospitality and travel funds that supported this work. K.S. thanks Stephen Randall for work done on the dWvH multiplet while at the University of Maryland. The authors   also     thank Konstantinos Koutrolikos for many helpful discussions throughout this work and S.J. Gates, Jr. for discussions and for providing the conceptual ideas that led to this~work.


\appendix

\section{Proof That \boldmath{\texorpdfstring{$t_{\rm IJ} = 1/2 \brV_{\rI\rJ}$}{tIJ=1VIJ}} Satisfies the \texorpdfstring{$so(N)$}{so(N)} Algebra}\label{a:soNProof}
Swapping $\brL_\rI$ with $\brR_\rI$ interchanges $\brV_{\rI\rJ}$ with $\brtV_{\rI\rJ}$, thus proving that $t=1/2 \brV_{\rI\rJ}$ satisfies the $so(N)$ algebra necessarily means that $t=1/2\brtV_{\rI\rJ}$ must also satisfy the $so(N)$ algebra. We therefore prove the latter, and the former follows by extension. Substituting $t = 1/2 \brV_{\rI\rJ}$ into the $so(N)$ algebra, Equation~(\ref{e:soN}), results in 
\begin{align}\label{e:VsoN}
[\brV_{\rm IJ} , \brV_{\rm KL}] ~=~ 2 i (\delta_{\rm I[L} \brV_{\rm K]J}-\delta_{\rm J[L} \brV_{\rm K]I}).
\end{align}

We now prove Equation~(\ref{e:VsoN}) using repeated use of the garden algebra, Equation~(\ref{e:GRdN}), rearranged as~follows
\begin{align}\label{e:GRdN2}
	\brL_{\rI} \brR_{\rJ} =~& 2 \delta_{\rI\rJ} -\brL_{\rJ} \brR_{\rI}.
\end{align}

\noindent We start by substituting the definition of $\brV_{\rI\rJ}$, Equation~(\ref{e:Vdef}), into the left hand side of Equation~(\ref{e:VsoN})
\begin{align}\label{e:VsoNProof1}
	[\brV_{\rm IJ} , \brV_{KL}] =~& \brV_{\rm IJ} \brV_{\rm KL} - \brV_{\rm KL} \brV_{\rm IJ} \cr
					=~& -\frac{1}{4} \left( \brL_{[\rI}\brR_{\rJ]}\brL_{[\rK}\brR_{\rL]} -  \brL_{[\rK}\brR_{\rL]}\brL_{[\rI}\brR_{\rJ]}   \right)
\end{align}

\noindent As an intermediate step, we make repeated use of Equation~(\ref{e:GRdN2}) to modify the last term, momentarily neglecting the antisymmetry between the indices $\rI$ and $\rJ$ and between $\rL$ and $\rK$
\begin{align}\label{e:VsoNProof2}
	 \brL_{\rK}\brR_{\rL}\brL_{\rI}\brR_{\rJ} =~& 2 \delta_{\rI\rL}\brL_\rK\brR_\rJ - \brL_\rK\brR_\rI\brL_\rL\brR_{\rJ} \cr
	 =~& 2 \delta_{\rI\rL}\brL_\rK\brR_\rJ - 2 \delta_{\rI\rK}\brL_\rL\brR_\rJ + \brL_\rI \brR_\rK\brL_\rL\brR_\rJ \cr
	 =~& 2 \delta_{\rI\rL}\brL_\rK\brR_\rJ - 2 \delta_{\rI\rK}\brL_\rL\brR_\rJ + 2 \delta_{\rJ\rL}\brL_\rI \brR_\rK - \brL_\rI \brR_\rK\brL_\rJ\brR_\rL  \cr
	  =~& 2 \delta_{\rI\rL}\brL_\rK\brR_\rJ - 2 \delta_{\rI\rK}\brL_\rL\brR_\rJ + 2 \delta_{\rJ\rL}\brL_\rI \brR_\rK - 2 \delta_{\rJ\rK}\brL_\rI \brR_\rL + \brL_\rI \brR_\rJ\brL_\rK\brR_\rL
\end{align}

\noindent Substituting this back into Equation~(\ref{e:VsoNProof1}) yields
\begin{align}
	[\brV_{\rm IJ} , \brV_{KL}] = - \frac{1}{2} &\left(\delta_{\rI[\rL}\brL_{\rK]}\brR_\rJ - \delta_{\rJ[\rL}\brL_{\rK]}\brR_\rI - \delta_{\rI[\rK}\brL_{\rL]}\brR_\rJ +\delta_{\rJ[\rK}\brL_{\rL]}\brR_\rI  \right. \cr
	&\left.-\delta_{\rI[\rL}\brL_{|\rJ|} \brR_{\rK]} +\delta_{\rJ[\rL}\brL_{|\rI|} \brR_{\rK]} -\delta_{\rI[\rK}\brL_{|\rJ|} \brR_{\rL]} -\delta_{\rJ[\rK}\brL_{|\rI|} \brR_{\rL]}  \right)
\end{align}

\noindent The first and fifth terms combine into a single term with $\brV_{\rm KJ}$, the second and sixth into $\brV_{\rm KI}$ and so on:
\begin{align}
 	[\brV_{\rm IJ} , \brV_{KL}] =~&~ i \left( \delta_{\rI[\rL}\brV_{\rK]\rJ} - \delta_{\rJ[\rL}\brV_{\rK]\rI} - \delta_{\rI[\rK}\brV_{\rL]\rJ} +\delta_{\rJ[\rK}\brV_{\rL]\rI} \right) \cr
 	=&~ 2 i \left( \delta_{\rI[\rL}\brV_{\rK]\rJ} - \delta_{\rJ[\rL}\brV_{\rK]\rI}\right)
\end{align}
QED
\section{Explicit \texorpdfstring{${\rm \textbf{L}}_{\rm I}$}{LI} and \texorpdfstring{${\rm \textbf{R}}_{\rm I}$}{RI} Matrices}\label{a:LRmatrices}
For the choice $c_0$ = 0, the explicit ${\rm \textbf{L}}_{\rm I}$  matrices for the dWvH multiplet are
\begin{align}
L_1 =~&
\text{\tiny $
	\left(
\begin{array}{cccccccccccccccccccc}
 1 & 0 & 0 & 0 & 0 & 0 & 0 & 0 & 0 & 0 & 0 & 0 & 0 & 1 & 0 & 0 & 0 & 0 & 0 & 0 \\
 0 & 0 & 0 & 0 & 1 & 0 & 0 & 0 & 0 & 0 & 0 & 0 & 0 & 0 & 0 & -1 & 0 & 0 & 0 & 0 \\
 0 & 0 & 0 & 0 & 0 & 0 & 0 & 0 & 1 & 0 & 0 & 0 & 1 & 0 & 0 & 0 & 0 & 0 & 0 & 0 \\
 1 & 0 & 0 & 0 & 0 & 0 & 1 & 0 & 0 & -1 & 0 & 0 & 0 & \frac{1}{2} & 0 & 0 & \frac{1}{2} & 0 & 0 & 0
   \\
 0 & 0 & 0 & 1 & 0 & 1 & 0 & 0 & 0 & 0 & 1 & 0 & 0 & 0 & \frac{1}{2} & 0 & 0 & 0 & 0 & \frac{1}{2}
   \\
 0 & 0 & 0 & 0 & 0 & 0 & 0 & 0 & 0 & 0 & 0 & 0 & 0 & 0 & -\frac{1}{2} & 0 & 0 & 0 & 0 & \frac{1}{2}
   \\
 0 & 1 & 0 & 0 & 0 & 0 & 0 & -1 & 1 & 0 & 0 & 0 & \frac{1}{2} & 0 & 0 & 0 & 0 & -\frac{1}{2} & 0 & 0
   \\
 0 & -1 & 0 & 0 & 0 & 0 & 0 & 0 & 0 & 0 & 0 & 0 & -\frac{1}{2} & 0 & 0 & 0 & 0 & \frac{1}{2} & 0 & 0
   \\
 0 & 0 & 0 & 0 & 0 & -1 & 0 & 0 & 0 & 0 & 0 & 0 & 0 & 0 & -\frac{1}{2} & 0 & 0 & 0 & 0 &
   -\frac{1}{2} \\
 0 & 0 & 0 & 0 & 0 & 0 & 0 & 0 & 0 & -1 & 0 & 0 & 0 & \frac{1}{2} & 0 & 0 & \frac{1}{2} & 0 & 0 & 0
   \\
 0 & 0 & -1 & 0 & 1 & 0 & 0 & 0 & 0 & 0 & 0 & 1 & 0 & 0 & 0 & -\frac{1}{2} & 0 & 0 & \frac{1}{2} & 0
   \\
 0 & 0 & 1 & 0 & 0 & 0 & 0 & 0 & 0 & 0 & 0 & 0 & 0 & 0 & 0 & \frac{1}{2} & 0 & 0 & -\frac{1}{2} & 0
   \\
 0 & 0 & 0 & 0 & 0 & 0 & 1 & 0 & 0 & 0 & 0 & 0 & 0 & \frac{1}{2} & 0 & 0 & \frac{1}{2} & 0 & 0 & 0
   \\
 0 & 0 & 0 & 0 & 0 & 0 & 0 & 0 & 0 & 0 & 1 & 0 & 0 & 0 & \frac{1}{2} & 0 & 0 & 0 & 0 & \frac{1}{2}
   \\
 1 & 0 & 0 & 0 & 0 & 0 & 0 & 0 & 0 & 0 & 0 & 0 & 0 & \frac{3}{2} & 0 & 0 & -\frac{1}{2} & 0 & 0 & 0
   \\
 0 & 0 & 0 & 0 & 1 & 0 & 0 & 0 & 0 & 0 & 0 & 0 & 0 & 0 & 0 & -\frac{3}{2} & 0 & 0 & -\frac{1}{2} & 0
   \\
 0 & 0 & 0 & 0 & 0 & 0 & 0 & 0 & 1 & 0 & 0 & 0 & \frac{3}{2} & 0 & 0 & 0 & 0 & \frac{1}{2} & 0 & 0
   \\
 0 & 0 & 0 & 0 & 0 & 0 & 0 & 0 & 0 & 0 & 0 & 1 & 0 & 0 & 0 & -\frac{1}{2} & 0 & 0 & \frac{1}{2} & 0
   \\
 0 & 0 & 0 & 1 & 0 & 0 & 0 & 0 & 0 & 0 & 0 & 0 & 0 & 0 & \frac{1}{2} & 0 & 0 & 0 & 0 & \frac{1}{2}
   \\
 0 & 0 & 0 & 0 & 0 & 0 & 0 & 1 & 0 & 0 & 0 & 0 & -\frac{1}{2} & 0 & 0 & 0 & 0 & \frac{1}{2} & 0 & 0
   \\
\end{array}
\right)
$
}
\end{align}

\begin{align}
L_2 =~&
\text{\tiny $
\left(
\begin{array}{cccccccccccccccccccc}
 0 & 1 & 0 & 0 & 0 & 0 & 0 & 0 & 0 & 0 & 0 & 0 & 1 & 0 & 0 & 0 & 0 & 0 & 0 & 0 \\
 0 & 0 & 0 & 0 & 0 & 1 & 0 & 0 & 0 & 0 & 0 & 0 & 0 & 0 & 1 & 0 & 0 & 0 & 0 & 0 \\
 0 & 0 & 0 & 0 & 0 & 0 & 0 & 0 & 0 & 1 & 0 & 0 & 0 & -1 & 0 & 0 & 0 & 0 & 0 & 0 \\
 0 & -1 & 0 & 0 & 0 & 0 & 0 & 1 & -1 & 0 & 0 & 0 & -\frac{1}{2} & 0 & 0 & 0 & 0 & \frac{1}{2} & 0 &
   0 \\
 0 & 0 & 1 & 0 & -1 & 0 & 0 & 0 & 0 & 0 & 0 & -1 & 0 & 0 & 0 & \frac{1}{2} & 0 & 0 & -\frac{1}{2} &
   0 \\
 0 & 0 & 0 & 0 & 0 & 0 & 0 & 0 & 0 & 0 & 0 & 0 & 0 & 0 & 0 & -\frac{1}{2} & 0 & 0 & -\frac{1}{2} & 0
   \\
 1 & 0 & 0 & 0 & 0 & 0 & 1 & 0 & 0 & -1 & 0 & 0 & 0 & \frac{1}{2} & 0 & 0 & \frac{1}{2} & 0 & 0 & 0
   \\
 1 & 0 & 0 & 0 & 0 & 0 & 0 & 0 & 0 & 0 & 0 & 0 & 0 & \frac{1}{2} & 0 & 0 & \frac{1}{2} & 0 & 0 & 0
   \\
 0 & 0 & 0 & 0 & 1 & 0 & 0 & 0 & 0 & 0 & 0 & 0 & 0 & 0 & 0 & -\frac{1}{2} & 0 & 0 & \frac{1}{2} & 0
   \\
 0 & 0 & 0 & 0 & 0 & 0 & 0 & 0 & 1 & 0 & 0 & 0 & \frac{1}{2} & 0 & 0 & 0 & 0 & -\frac{1}{2} & 0 & 0
   \\
 0 & 0 & 0 & 1 & 0 & 1 & 0 & 0 & 0 & 0 & 1 & 0 & 0 & 0 & \frac{1}{2} & 0 & 0 & 0 & 0 & \frac{1}{2}
   \\
 0 & 0 & 0 & 1 & 0 & 0 & 0 & 0 & 0 & 0 & 0 & 0 & 0 & 0 & \frac{1}{2} & 0 & 0 & 0 & 0 & \frac{1}{2}
   \\
 0 & 0 & 0 & 0 & 0 & 0 & 0 & 1 & 0 & 0 & 0 & 0 & -\frac{1}{2} & 0 & 0 & 0 & 0 & \frac{1}{2} & 0 & 0
   \\
 0 & 0 & 0 & 0 & 0 & 0 & 0 & 0 & 0 & 0 & 0 & 1 & 0 & 0 & 0 & -\frac{1}{2} & 0 & 0 & \frac{1}{2} & 0
   \\
 0 & 1 & 0 & 0 & 0 & 0 & 0 & 0 & 0 & 0 & 0 & 0 & \frac{3}{2} & 0 & 0 & 0 & 0 & \frac{1}{2} & 0 & 0
   \\
 0 & 0 & 0 & 0 & 0 & 1 & 0 & 0 & 0 & 0 & 0 & 0 & 0 & 0 & \frac{3}{2} & 0 & 0 & 0 & 0 & -\frac{1}{2}
   \\
 0 & 0 & 0 & 0 & 0 & 0 & 0 & 0 & 0 & 1 & 0 & 0 & 0 & -\frac{3}{2} & 0 & 0 & \frac{1}{2} & 0 & 0 & 0
   \\
 0 & 0 & 0 & 0 & 0 & 0 & 0 & 0 & 0 & 0 & -1 & 0 & 0 & 0 & -\frac{1}{2} & 0 & 0 & 0 & 0 &
   -\frac{1}{2} \\
 0 & 0 & -1 & 0 & 0 & 0 & 0 & 0 & 0 & 0 & 0 & 0 & 0 & 0 & 0 & -\frac{1}{2} & 0 & 0 & \frac{1}{2} & 0
   \\
 0 & 0 & 0 & 0 & 0 & 0 & -1 & 0 & 0 & 0 & 0 & 0 & 0 & -\frac{1}{2} & 0 & 0 & -\frac{1}{2} & 0 & 0 &
   0 \\
\end{array}
\right)
$
}
\end{align}

\begin{align}
L_3 =~&
\text{\tiny $
\left(
\begin{array}{cccccccccccccccccccc}
 0 & 0 & 1 & 0 & 0 & 0 & 0 & 0 & 0 & 0 & 0 & 0 & 0 & 0 & 0 & 1 & 0 & 0 & 0 & 0 \\
 0 & 0 & 0 & 0 & 0 & 0 & 1 & 0 & 0 & 0 & 0 & 0 & 0 & 1 & 0 & 0 & 0 & 0 & 0 & 0 \\
 0 & 0 & 0 & 0 & 0 & 0 & 0 & 0 & 0 & 0 & 1 & 0 & 0 & 0 & 1 & 0 & 0 & 0 & 0 & 0 \\
 0 & 0 & -1 & 0 & 1 & 0 & 0 & 0 & 0 & 0 & 0 & 1 & 0 & 0 & 0 & -\frac{1}{2} & 0 & 0 & \frac{1}{2} & 0
   \\
 0 & -1 & 0 & 0 & 0 & 0 & 0 & 1 & -1 & 0 & 0 & 0 & -\frac{1}{2} & 0 & 0 & 0 & 0 & \frac{1}{2} & 0 &
   0 \\
 0 & 0 & 0 & 0 & 0 & 0 & 0 & 0 & 0 & 0 & 0 & 0 & \frac{1}{2} & 0 & 0 & 0 & 0 & \frac{1}{2} & 0 & 0
   \\
 0 & 0 & 0 & 1 & 0 & 1 & 0 & 0 & 0 & 0 & 1 & 0 & 0 & 0 & \frac{1}{2} & 0 & 0 & 0 & 0 & \frac{1}{2}
   \\
 0 & 0 & 0 & 1 & 0 & 0 & 0 & 0 & 0 & 0 & 0 & 0 & 0 & 0 & \frac{1}{2} & 0 & 0 & 0 & 0 & \frac{1}{2}
   \\
 0 & 0 & 0 & 0 & 0 & 0 & 0 & 1 & 0 & 0 & 0 & 0 & -\frac{1}{2} & 0 & 0 & 0 & 0 & \frac{1}{2} & 0 & 0
   \\
 0 & 0 & 0 & 0 & 0 & 0 & 0 & 0 & 0 & 0 & 0 & 1 & 0 & 0 & 0 & -\frac{1}{2} & 0 & 0 & \frac{1}{2} & 0
   \\
 -1 & 0 & 0 & 0 & 0 & 0 & -1 & 0 & 0 & 1 & 0 & 0 & 0 & -\frac{1}{2} & 0 & 0 & -\frac{1}{2} & 0 & 0 &
   0 \\
 -1 & 0 & 0 & 0 & 0 & 0 & 0 & 0 & 0 & 0 & 0 & 0 & 0 & -\frac{1}{2} & 0 & 0 & -\frac{1}{2} & 0 & 0 &
   0 \\
 0 & 0 & 0 & 0 & -1 & 0 & 0 & 0 & 0 & 0 & 0 & 0 & 0 & 0 & 0 & \frac{1}{2} & 0 & 0 & -\frac{1}{2} & 0
   \\
 0 & 0 & 0 & 0 & 0 & 0 & 0 & 0 & -1 & 0 & 0 & 0 & -\frac{1}{2} & 0 & 0 & 0 & 0 & \frac{1}{2} & 0 & 0
   \\
 0 & 0 & 1 & 0 & 0 & 0 & 0 & 0 & 0 & 0 & 0 & 0 & 0 & 0 & 0 & \frac{3}{2} & 0 & 0 & \frac{1}{2} & 0
   \\
 0 & 0 & 0 & 0 & 0 & 0 & 1 & 0 & 0 & 0 & 0 & 0 & 0 & \frac{3}{2} & 0 & 0 & -\frac{1}{2} & 0 & 0 & 0
   \\
 0 & 0 & 0 & 0 & 0 & 0 & 0 & 0 & 0 & 0 & 1 & 0 & 0 & 0 & \frac{3}{2} & 0 & 0 & 0 & 0 & -\frac{1}{2}
   \\
 0 & 0 & 0 & 0 & 0 & 0 & 0 & 0 & 0 & 1 & 0 & 0 & 0 & -\frac{1}{2} & 0 & 0 & -\frac{1}{2} & 0 & 0 & 0
   \\
 0 & 1 & 0 & 0 & 0 & 0 & 0 & 0 & 0 & 0 & 0 & 0 & \frac{1}{2} & 0 & 0 & 0 & 0 & -\frac{1}{2} & 0 & 0
   \\
 0 & 0 & 0 & 0 & 0 & 1 & 0 & 0 & 0 & 0 & 0 & 0 & 0 & 0 & \frac{1}{2} & 0 & 0 & 0 & 0 & \frac{1}{2}
   \\
\end{array}
\right)
$
}
\end{align}

\begin{align}
L_4 =~&
\text{\tiny $
\left(
\begin{array}{cccccccccccccccccccc}
 0 & 0 & 0 & 1 & 0 & 0 & 0 & 0 & 0 & 0 & 0 & 0 & 0 & 0 & 1 & 0 & 0 & 0 & 0 & 0 \\
 0 & 0 & 0 & 0 & 0 & 0 & 0 & 1 & 0 & 0 & 0 & 0 & -1 & 0 & 0 & 0 & 0 & 0 & 0 & 0 \\
 0 & 0 & 0 & 0 & 0 & 0 & 0 & 0 & 0 & 0 & 0 & 1 & 0 & 0 & 0 & -1 & 0 & 0 & 0 & 0 \\
 0 & 0 & 0 & 1 & 0 & 1 & 0 & 0 & 0 & 0 & 1 & 0 & 0 & 0 & \frac{1}{2} & 0 & 0 & 0 & 0 & \frac{1}{2}
   \\
 -1 & 0 & 0 & 0 & 0 & 0 & -1 & 0 & 0 & 1 & 0 & 0 & 0 & -\frac{1}{2} & 0 & 0 & -\frac{1}{2} & 0 & 0 &
   0 \\
 0 & 0 & 0 & 0 & 0 & 0 & 0 & 0 & 0 & 0 & 0 & 0 & 0 & \frac{1}{2} & 0 & 0 & -\frac{1}{2} & 0 & 0 & 0
   \\
 0 & 0 & 1 & 0 & -1 & 0 & 0 & 0 & 0 & 0 & 0 & -1 & 0 & 0 & 0 & \frac{1}{2} & 0 & 0 & -\frac{1}{2} &
   0 \\
 0 & 0 & -1 & 0 & 0 & 0 & 0 & 0 & 0 & 0 & 0 & 0 & 0 & 0 & 0 & -\frac{1}{2} & 0 & 0 & \frac{1}{2} & 0
   \\
 0 & 0 & 0 & 0 & 0 & 0 & -1 & 0 & 0 & 0 & 0 & 0 & 0 & -\frac{1}{2} & 0 & 0 & -\frac{1}{2} & 0 & 0 &
   0 \\
 0 & 0 & 0 & 0 & 0 & 0 & 0 & 0 & 0 & 0 & -1 & 0 & 0 & 0 & -\frac{1}{2} & 0 & 0 & 0 & 0 &
   -\frac{1}{2} \\
 0 & 1 & 0 & 0 & 0 & 0 & 0 & -1 & 1 & 0 & 0 & 0 & \frac{1}{2} & 0 & 0 & 0 & 0 & -\frac{1}{2} & 0 & 0
   \\
 0 & -1 & 0 & 0 & 0 & 0 & 0 & 0 & 0 & 0 & 0 & 0 & -\frac{1}{2} & 0 & 0 & 0 & 0 & \frac{1}{2} & 0 & 0
   \\
 0 & 0 & 0 & 0 & 0 & -1 & 0 & 0 & 0 & 0 & 0 & 0 & 0 & 0 & -\frac{1}{2} & 0 & 0 & 0 & 0 &
   -\frac{1}{2} \\
 0 & 0 & 0 & 0 & 0 & 0 & 0 & 0 & 0 & -1 & 0 & 0 & 0 & \frac{1}{2} & 0 & 0 & \frac{1}{2} & 0 & 0 & 0
   \\
 0 & 0 & 0 & 1 & 0 & 0 & 0 & 0 & 0 & 0 & 0 & 0 & 0 & 0 & \frac{3}{2} & 0 & 0 & 0 & 0 & -\frac{1}{2}
   \\
 0 & 0 & 0 & 0 & 0 & 0 & 0 & 1 & 0 & 0 & 0 & 0 & -\frac{3}{2} & 0 & 0 & 0 & 0 & -\frac{1}{2} & 0 & 0
   \\
 0 & 0 & 0 & 0 & 0 & 0 & 0 & 0 & 0 & 0 & 0 & 1 & 0 & 0 & 0 & -\frac{3}{2} & 0 & 0 & -\frac{1}{2} & 0
   \\
 0 & 0 & 0 & 0 & 0 & 0 & 0 & 0 & -1 & 0 & 0 & 0 & -\frac{1}{2} & 0 & 0 & 0 & 0 & \frac{1}{2} & 0 & 0
   \\
 -1 & 0 & 0 & 0 & 0 & 0 & 0 & 0 & 0 & 0 & 0 & 0 & 0 & -\frac{1}{2} & 0 & 0 & -\frac{1}{2} & 0 & 0 &
   0 \\
 0 & 0 & 0 & 0 & -1 & 0 & 0 & 0 & 0 & 0 & 0 & 0 & 0 & 0 & 0 & \frac{1}{2} & 0 & 0 & -\frac{1}{2} & 0
   \\
\end{array}
\right)
$
}
\end{align}

The ${\rm \textbf{R}}_{\rm I}$  matrices are inverses of the ${\rm \textbf{L}}_{\rm I}$  matrices: ${\rm \textbf{R}}_{\rm I}  = {\rm \textbf{L}}_{\rm I}^{-1}$.

\newpage

\section{Explicit Form for the \texorpdfstring{$\brtV_{\rm IJ}$}{VIJ} Matrices for the dWVH Multiplet in a \boldmath{\texorpdfstring{$20 \times 20$}{20x20}} Tensor Product Basis}\label{a:tildeVmatrices}
It would be instructive to construct a tensor product basis into which $20 \times 20$ matrices can be displayed. This is particularly useful for the $\tilde{V}_{\rm IJ}$ matrices, as shown in this section. In~\cite{G-1}, a~\mbox{$so(4) = su(2) \times su(2)$} basis of $4 \times 4$ matrices is defined to illustrate how the fundamental adinkras CM, VM, and TM possess this symmetry even at the one-dimensional adinkra level.  This $so(4)$ basis is 
\label{a:ab}
\begin{align*}
	\alpha_1 =&  \left(
\begin{array}{cccc}
 0 & 0 & 0 & -i \\
 0 & 0 & -i & 0 \\
 0 & i & 0 & 0 \\
 i & 0 & 0 & 0 \\
\end{array}
\right),~~~ 
	\alpha_2 =  \left(
\begin{array}{cccc}
 0 & -i & 0 & 0 \\
 i & 0 & 0 & 0 \\
 0 & 0 & 0 & -i \\
 0 & 0 & i & 0 \\
\end{array}
\right),~~~ 
	\alpha_3 = \left(
\begin{array}{cccc}
 0 & 0 & -i & 0 \\
 0 & 0 & 0 & i \\
 i & 0 & 0 & 0 \\
 0 & -i & 0 & 0 \\
\end{array}
\right) 
\end{align*}

\begin{align*}
	\beta_1 = &  \left(
\begin{array}{cccc}
 0 & 0 & 0 & -i \\
 0 & 0 & i & 0 \\
 0 & -i & 0 & 0 \\
 i & 0 & 0 & 0 \\
\end{array}
\right),~~~ 
	\beta_2 =  \left(
\begin{array}{cccc}
 0 & 0 & -i & 0 \\
 0 & 0 & 0 & -i \\
 i & 0 & 0 & 0 \\
 0 & i & 0 & 0 \\
\end{array}
\right) ,~~~ 
	\beta_3 =  \left(
\begin{array}{cccc}
 0 & -i & 0 & 0 \\
 i & 0 & 0 & 0 \\
 0 & 0 & 0 & i \\
 0 & 0 & -i & 0 \\
\end{array}
\right).
\end{align*}

These $\beta_{\hat{a}}$ matrices are not to be confused with the auxiliary fermion $\beta_a$ for \nmSG.
In terms of tensor products of Pauli spin matrices $\sigma^i$ and the $2\times2$ identity matrix $I_2$, this can be written as
\begin{align*}
	\alpha_1 =& \sigma^2 \otimes \sigma^1,~~~\alpha_2 = \rI_2 \otimes \sigma^2,~~~\alpha_3 = \sigma^2 \otimes \sigma^3 \\
	\beta_1 =& \sigma^1 \otimes \sigma^2,~~~\beta_2 = \sigma^2 \otimes \rI_2,~~~\beta_3 = \sigma^3 \otimes \sigma^2 .
\end{align*}
Augmenting these six matrices with the $4 \times 4$ identity $\rI_4$, we construct a sixteen element $l(4,R)$ basis as follows
\begin{align}
	\rI_4,~~~\alpha_{\hat{a}},~~~ \beta_{\hat{a}},~~~ \alpha_{\hat{a}}\beta_{\hat{b}}.
\end{align}
Next, we introduce an $l(5,R)$ basis of $5 \times 5$ matrices:
\begin{align}
	\omega_0^5 =& \frac{1}{10}\rI_5,~~~
	\omega_1^5 = \frac{1}{4} 
	\text{\scriptsize $\left(
\begin{array}{ccccc}
 0 & 1 & 0 & 0 & 0 \\
 1 & 0 & 0 & 0 & 0 \\
 0 & 0 & 0 & 0 & 0 \\
 0 & 0 & 0 & 0 & 0 \\
 0 & 0 & 0 & 0 & 0 \\
\end{array}
\right)
$},~~~
	\omega_2^5 = \frac{1}{4}
	\text{\scriptsize $\left(
\begin{array}{ccccc}
 0 & -1 & 0 & 0 & 0 \\
 1 & 0 & 0 & 0 & 0 \\
 0 & 0 & 0 & 0 & 0 \\
 0 & 0 & 0 & 0 & 0 \\
 0 & 0 & 0 & 0 & 0 \\
\end{array}
\right)
$}\cr
	\omega_3^5 =& \frac{1}{4}
	\text{\scriptsize $\left(
\begin{array}{ccccc}
 1 & 0 & 0 & 0 & 0 \\
 0 & -1 & 0 & 0 & 0 \\
 0 & 0 & 0 & 0 & 0 \\
 0 & 0 & 0 & 0 & 0 \\
 0 & 0 & 0 & 0 & 0 \\
\end{array}
\right)
$},~~~
	\omega_4^5 = \frac{1}{4}
	\text{\scriptsize $\left(
\begin{array}{ccccc}
 0 & 0 & 1 & 0 & 0 \\
 0 & 0 & 0 & 0 & 0 \\
 1 & 0 & 0 & 0 & 0 \\
 0 & 0 & 0 & 0 & 0 \\
 0 & 0 & 0 & 0 & 0 \\
\end{array}
\right)
$},~~~
	\omega_5^5 = \frac{1}{4}
	\text{\scriptsize $\left(
\begin{array}{ccccc}
 0 & 0 & -1 & 0 & 0 \\
 0 & 0 & 0 & 0 & 0 \\
 1 & 0 & 0 & 0 & 0 \\
 0 & 0 & 0 & 0 & 0 \\
 0 & 0 & 0 & 0 & 0 \\
\end{array}
\right)
$}\cr
	\omega_6^5 =& \frac{1}{4}
	\text{\scriptsize $\left(
\begin{array}{ccccc}
 0 & 0 & 0 & 0 & 0 \\
 0 & 0 & 1 & 0 & 0 \\
 0 & 1 & 0 & 0 & 0 \\
 0 & 0 & 0 & 0 & 0 \\
 0 & 0 & 0 & 0 & 0 \\
\end{array}
\right)
$},~~~
	\omega_7^5 = \frac{1}{4}
	\text{\scriptsize $\left(
\begin{array}{ccccc}
 0 & 0 & 0 & 0 & 0 \\
 0 & 0 & -1 & 0 & 0 \\
 0 & 1 & 0 & 0 & 0 \\
 0 & 0 & 0 & 0 & 0 \\
 0 & 0 & 0 & 0 & 0 \\
\end{array}
\right)
$},~~~
	\omega_8^5 = \frac{1}{12}
	\text{\scriptsize $\left(
\begin{array}{ccccc}
 1 & 0 & 0 & 0 & 0 \\
 0 & 1 & 0 & 0 & 0 \\
 0 & 0 & -2 & 0 & 0 \\
 0 & 0 & 0 & 0 & 0 \\
 0 & 0 & 0 & 0 & 0 \\
\end{array}
\right)
$}\cr
	\omega_9^5 =& \frac{1}{4}
	\text{\scriptsize $\left(
\begin{array}{ccccc}
 0 & 0 & 0 & 1 & 0 \\
 0 & 0 & 0 & 0 & 0 \\
 0 & 0 & 0 & 0 & 0 \\
 1 & 0 & 0 & 0 & 0 \\
 0 & 0 & 0 & 0 & 0 \\
\end{array}
\right)
$},~~~
	\omega_{10}^5 = \frac{1}{4}
	\text{\scriptsize $\left(
\begin{array}{ccccc}
 0 & 0 & 0 & -1 & 0 \\
 0 & 0 & 0 & 0 & 0 \\
 0 & 0 & 0 & 0 & 0 \\
 1 & 0 & 0 & 0 & 0 \\
 0 & 0 & 0 & 0 & 0 \\
\end{array}
\right)
$},~~~
	\omega_{11}^5 = \frac{1}{4}
	\text{\scriptsize $\left(
\begin{array}{ccccc}
 0 & 0 & 0 & 0 & 0 \\
 0 & 0 & 0 & 1 & 0 \\
 0 & 0 & 0 & 0 & 0 \\
 0 & 1 & 0 & 0 & 0 \\
 0 & 0 & 0 & 0 & 0 \\
\end{array}
\right)
$}\cr
	\omega_{12}^5 =& \frac{1}{4}
	\text{\scriptsize $\left(
\begin{array}{ccccc}
 0 & 0 & 0 & 0 & 0 \\
 0 & 0 & 0 & -1 & 0 \\
 0 & 0 & 0 & 0 & 0 \\
 0 & 1 & 0 & 0 & 0 \\
 0 & 0 & 0 & 0 & 0 \\
\end{array}
\right)
$},~~~
	\omega_{13}^5 = \frac{1}{4}
	\text{\scriptsize $\left(
\begin{array}{ccccc}
 0 & 0 & 0 & 0 & 0 \\
 0 & 0 & 0 & 0 & 0 \\
 0 & 0 & 0 & 1 & 0 \\
 0 & 0 & 1 & 0 & 0 \\
 0 & 0 & 0 & 0 & 0 \\
\end{array}
\right)
$},~~~
	\omega_{14}^5 = \frac{1}{4}
	\text{\scriptsize $\left(
\begin{array}{ccccc}
 0 & 0 & 0 & 0 & 0 \\
 0 & 0 & 0 & 0 & 0 \\
 0 & 0 & 0 & -1 & 0 \\
 0 & 0 & 1 & 0 & 0 \\
 0 & 0 & 0 & 0 & 0 \\
\end{array}
\right)
$}\cr
	\omega_{15}^5 =& \frac{1}{24}
	\text{\scriptsize $\left(
\begin{array}{ccccc}
1 & 0 & 0 & 0 & 0 \\
 0 & 1 & 0 & 0 & 0 \\
 0 & 0 & 1 & 0 & 0 \\
 0 & 0 & 0 & -3 & 0 \\
 0 & 0 & 0 & 0 & 0 \\
\end{array}
\right)
$},~~~
	\omega_{16}^5 = \frac{1}{4}
	\text{\scriptsize $\left(
\begin{array}{ccccc}
 0 & 0 & 0 & 0 & 1 \\
 0 & 0 & 0 & 0 & 0 \\
 0 & 0 & 0 & 0 & 0 \\
 0 & 0 & 0 & 0 & 0 \\
 1 & 0 & 0 & 0 & 0 \\
\end{array}
\right)
$},~~~
	\omega_{17}^5 = \frac{1}{4}
	\text{\scriptsize $\left(
\begin{array}{ccccc}
 0 & 0 & 0 & 0 & -1 \\
 0 & 0 & 0 & 0 & 0 \\
 0 & 0 & 0 & 0 & 0 \\
 0 & 0 & 0 & 0 & 0 \\
 1 & 0 & 0 & 0 & 0 \\
\end{array}
\right)
$}\cr
	\omega_{18}^5 =& \frac{1}{4}
	\text{\scriptsize $\left(
\begin{array}{ccccc}
 0 & 0 & 0 & 0 & 0 \\
 0 & 0 & 0 & 0 & 1 \\
 0 & 0 & 0 & 0 & 0 \\
 0 & 0 & 0 & 0 & 0 \\
 0 & 1 & 0 & 0 & 0 \\
\end{array}
\right)
$},~~~
	\omega_{19}^5 = \frac{1}{4}
	\text{\scriptsize $\left(
\begin{array}{ccccc}
 0 & 0 & 0 & 0 & 0 \\
 0 & 0 & 0 & 0 & -1 \\
 0 & 0 & 0 & 0 & 0 \\
 0 & 0 & 0 & 0 & 0 \\
 0 & 1 & 0 & 0 & 0 \\
\end{array}
\right)
$},~~~
	\omega_{20}^5 = \frac{1}{4}
	\text{\scriptsize $\left(
\begin{array}{ccccc}
 0 & 0 & 0 & 0 & 0 \\
 0 & 0 & 0 & 0 & 0 \\
 0 & 0 & 0 & 0 & 1 \\
 0 & 0 & 0 & 0 & 0 \\
 0 & 0 & 1 & 0 & 0 \\
\end{array}
\right)
$}\cr
	\omega_{21}^5 =& \frac{1}{4}
	\text{\scriptsize $\left(
\begin{array}{ccccc}
 0 & 0 & 0 & 0 & 0 \\
 0 & 0 & 0 & 0 & 0 \\
 0 & 0 & 0 & 0 & -1 \\
 0 & 0 & 0 & 0 & 0 \\
 0 & 0 & 1 & 0 & 0 \\
\end{array}
\right)
$},~~~
	\omega_{22}^5 = \frac{1}{4}
	\text{\scriptsize $\left(
\begin{array}{ccccc}
 0 & 0 & 0 & 0 & 0 \\
 0 & 0 & 0 & 0 & 0 \\
 0 & 0 & 0 & 0 & 0 \\
 0 & 0 & 0 & 0 & 1 \\
 0 & 0 & 0 & 1 & 0 \\
\end{array}
\right)
$} ,~~~
	\omega_{23}^5 = \frac{1}{4}
	\text{\scriptsize $\left(
\begin{array}{ccccc}
 0 & 0 & 0 & 0 & 0 \\
 0 & 0 & 0 & 0 & 0 \\
 0 & 0 & 0 & 0 & 0 \\
 0 & 0 & 0 & 0 & -1 \\
 0 & 0 & 0 & 1 & 0 \\
\end{array}
\right)
$}\cr
	\omega_{24}^5 =& \frac{1}{40}
	\text{\scriptsize $\left(
\begin{array}{ccccc}
 1 & 0 & 0 & 0 & 0 \\
 0 & 1 & 0 & 0 & 0 \\
 0 & 0 & 1 & 0 & 0 \\
 0 & 0 & 0 & 1 & 0 \\
 0 & 0 & 0 & 0 & -4 \\
\end{array}
\right)
$}
\end{align}

The above basis corresponds to the normalization choice nz[5] = 2 in the data file \emph{dWvH.m} found at a GitHub \href{https://hepthools.github.io/Data/}{data repository}, although the user can choose any normalization for any representation. For~instance, the form of the first three  $l(5,R)$ matrices take the general form
\begin{align}
	\omega_0^5 =& \frac{1}{5 nz[5]}\rI_5,~~~
	\omega_1^5 = \frac{1}{2 nz[5]} 
	\text{\tiny $\left(
\begin{array}{ccccc}
 0 & 1 & 0 & 0 & 0 \\
 1 & 0 & 0 & 0 & 0 \\
 0 & 0 & 0 & 0 & 0 \\
 0 & 0 & 0 & 0 & 0 \\
 0 & 0 & 0 & 0 & 0 \\
\end{array}
\right)
$},~~~
	\omega_2^5 = \frac{1}{2 nz[5]}
	\text{\scriptsize $\left(
\begin{array}{ccccc}
 0 & -1 & 0 & 0 & 0 \\
 1 & 0 & 0 & 0 & 0 \\
 0 & 0 & 0 & 0 & 0 \\
 0 & 0 & 0 & 0 & 0 \\
 0 & 0 & 0 & 0 & 0 \\
\end{array}
\right)
$}
\end{align}

\noindent The other $l(5,R)$ matrices have similar generalized normalizations.

For the choice $c_0$ = 0 and normalization nz[5] = 2, the explicit form for the $\brtV_{\rm IJ}$  matrices for the dWvH multiplet are
\begin{align}
	\brtV_{\rm 12} = & ~2 i \omega _9^5\otimes \left(\alpha _3 \beta _2\right)+2 i \omega _{10}^5\otimes \left(\alpha _3
   \beta _2\right)-2 i \omega _{13}^5\otimes \left(\alpha _1 \beta _2\right)-2 i \omega
   _{14}^5\otimes \left(\alpha _1 \beta _2\right)+2 i \omega _{16}^5\otimes \left(\alpha _3 \beta
   _1\right) \cr
   &+2 i \omega _{17}^5\otimes \left(\alpha _3 \beta _1\right)-2 i \omega _{20}^5\otimes
   \left(\alpha _1 \beta _1\right)-2 i \omega _{21}^5\otimes \left(\alpha _1 \beta _1\right)-2 i
   \omega _{23}^5\otimes \left(\alpha _2 \beta _3\right)+2 \omega _1^5\otimes \alpha _1 \cr
   &+2 \omega
   _2^5\otimes \alpha _1-2 \omega _3^5\otimes \alpha _2+2 \omega _6^5\otimes \alpha _3-2 \omega
   _7^5\otimes \alpha _3+2 \omega _8^5\otimes \alpha _2+5 \omega _{15}^5\otimes \alpha _2+5 \omega
   _{24}^5\otimes \alpha _2 \cr
   &-2 \omega _0^5\otimes \beta _3+2 \omega _{11}^5\otimes \beta _2-2 \omega
   _{12}^5\otimes \beta _2+3 \omega _{15}^5\otimes \beta _3+2 \omega _{18}^5\otimes \beta _1-2
   \omega _{19}^5\otimes \beta _1 \cr
   &+3 \omega _{24}^5\otimes \beta _3-4 i \omega _5^5\otimes
   \text{I}_4+2 i \omega _{23}^5\otimes \text{I}_4
\end{align}
\begin{align}
	\brtV_{\rm 13} = & -2 i \omega _9^5\otimes \left(\alpha _2 \beta _2\right)-2 i \omega _{10}^5\otimes \left(\alpha _2
   \beta _2\right)+2 i \omega _{11}^5\otimes \left(\alpha _1 \beta _2\right)+2 i \omega
   _{12}^5\otimes \left(\alpha _1 \beta _2\right)\cr
   &-2 i \omega _{16}^5\otimes \left(\alpha _2 \beta
   _1\right) -2 i \omega _{17}^5\otimes \left(\alpha _2 \beta _1\right)+2 i \omega _{18}^5\otimes
   \left(\alpha _1 \beta _1\right)+2 i \omega _{19}^5\otimes \left(\alpha _1 \beta _1\right)-2 i
   \omega _{23}^5\otimes \left(\alpha _3 \beta _3\right)\cr
   &+2 \omega _4^5\otimes \alpha _1 +2 \omega
   _5^5\otimes \alpha _1+2 \omega _6^5\otimes \alpha _2+2 \omega _7^5\otimes \alpha _2-4 \omega
   _8^5\otimes \alpha _3\cr
   &+5 \omega _{15}^5\otimes \alpha _3+5 \omega _{24}^5\otimes \alpha _3+2
   \omega _{13}^5\otimes \beta _2 -2 \omega _{14}^5\otimes \beta _2-3 \omega _{15}^5\otimes \beta
   _2+2 \omega _{20}^5\otimes \beta _1-2 \omega _{21}^5\otimes \beta _1 \cr
   &-2 \omega _{22}^5\otimes
   \beta _1+5 \omega _{24}^5\otimes \beta _2+4 i \omega _2^5\otimes \text{I}_4 
\end{align}
\begin{align}
	\brtV_{\rm 14} = &  -2 i \omega _{11}^5\otimes \left(\alpha _3 \beta _2\right)-2 i \omega _{12}^5\otimes \left(\alpha _3
   \beta _2\right) +2 i \omega _{13}^5\otimes \left(\alpha _2 \beta _2\right)+2 i \omega
   _{14}^5\otimes \left(\alpha _2 \beta _2\right)\cr
   &-2 i \omega _{18}^5\otimes \left(\alpha _3 \beta
   _1\right)-2 i \omega _{19}^5\otimes \left(\alpha _3 \beta _1\right)+2 i \omega _{20}^5\otimes
   \left(\alpha _2 \beta _1\right)+2 i \omega _{21}^5\otimes \left(\alpha _2 \beta _1\right)-2 i
   \omega _{23}^5\otimes \left(\alpha _1 \beta _3\right)\cr
   &+2 \omega _1^5\otimes \alpha _2-2 \omega
   _2^5\otimes \alpha _2+2 \omega _3^5\otimes \alpha _1+2 \omega _4^5\otimes \alpha _3-2 \omega
   _5^5\otimes \alpha _3+2 \omega _8^5\otimes \alpha _1+5 \omega _{15}^5\otimes \alpha _1\cr
   &+5 \omega
   _{24}^5\otimes \alpha _1+2 \omega _9^5\otimes \beta _2-2 \omega _{10}^5\otimes \beta _2+3 \omega
   _{15}^5\otimes \beta _1+2 \omega _{16}^5\otimes \beta _1-2 \omega _{17}^5\otimes \beta _1-2
   \omega _{22}^5\otimes \beta _2\cr
   &-5 \omega _{24}^5\otimes \beta _1+4 i \omega _7^5\otimes \text{I}_4
\end{align}
\begin{align}
	\brtV_{\rm 23} = & -2 i \omega _{11}^5\otimes \left(\alpha _3 \beta _2\right)-2 i \omega _{12}^5\otimes \left(\alpha _3
   \beta _2\right)+2 i \omega _{13}^5\otimes \left(\alpha _2 \beta _2\right)+2 i \omega
   _{14}^5\otimes \left(\alpha _2 \beta _2\right)\cr
   &-2 i \omega _{18}^5\otimes \left(\alpha _3 \beta
   _1\right)-2 i \omega _{19}^5\otimes \left(\alpha _3 \beta _1\right)+2 i \omega _{20}^5\otimes
   \left(\alpha _2 \beta _1\right)+2 i \omega _{21}^5\otimes \left(\alpha _2 \beta _1\right)\cr
   &-2 i
   \omega _{23}^5\otimes \left(\alpha _1 \beta _3\right)+2 \omega _1^5\otimes \alpha _2-2 \omega
   _2^5\otimes \alpha _2+2 \omega _3^5\otimes \alpha _1+2 \omega _4^5\otimes \alpha _3-2 \omega
   _5^5\otimes \alpha _3\cr
   &+2 \omega _8^5\otimes \alpha _1+5 \omega _{15}^5\otimes \alpha _1+5 \omega
   _{24}^5\otimes \alpha _1+2 \omega _9^5\otimes \beta _2-2 \omega _{10}^5\otimes \beta _2-3 \omega
   _{15}^5\otimes \beta _1\cr
   &+2 \omega _{16}^5\otimes \beta _1-2 \omega _{17}^5\otimes \beta _1+2
   \omega _{22}^5\otimes \beta _2+5 \omega _{24}^5\otimes \beta _1+4 i \omega _7^5\otimes \text{I}_4
\end{align}
\begin{align}
	\brtV_{\rm 24} = & ~2 i \omega _9^5\otimes \left(\alpha _2 \beta _2\right)+2 i \omega _{10}^5\otimes \left(\alpha _2
   \beta _2\right)-2 i \omega _{11}^5\otimes \left(\alpha _1 \beta _2\right)-2 i \omega
   _{12}^5\otimes \left(\alpha _1 \beta _2\right)+2 i \omega _{16}^5\otimes \left(\alpha _2 \beta
   _1\right)\cr
   &+2 i \omega _{17}^5\otimes \left(\alpha _2 \beta _1\right)-2 i \omega _{18}^5\otimes
   \left(\alpha _1 \beta _1\right)-2 i \omega _{19}^5\otimes \left(\alpha _1 \beta _1\right)+2 i
   \omega _{23}^5\otimes \left(\alpha _3 \beta _3\right)-2 \omega _4^5\otimes \alpha _1\cr
   &-2 \omega
   _5^5\otimes \alpha _1-2 \omega _6^5\otimes \alpha _2-2 \omega _7^5\otimes \alpha _2+4 \omega
   _8^5\otimes \alpha _3-5 \omega _{15}^5\otimes \alpha _3-5 \omega _{24}^5\otimes \alpha _3-2
   \omega _{13}^5\otimes \beta _2\cr
   &+2 \omega _{14}^5\otimes \beta _2-3 \omega _{15}^5\otimes \beta
   _2-2 \omega _{20}^5\otimes \beta _1+2 \omega _{21}^5\otimes \beta _1-2 \omega _{22}^5\otimes
   \beta _1+5 \omega _{24}^5\otimes \beta _2\cr
   &-4 i \omega _2^5\otimes \text{I}_4
\end{align}
\begin{align}
	\brtV_{34} = & ~2 i \omega _9^5\otimes \left(\alpha _3 \beta _2\right)+2 i \omega _{10}^5\otimes \left(\alpha _3
   \beta _2\right)-2 i \omega _{13}^5\otimes \left(\alpha _1 \beta _2\right)-2 i \omega
   _{14}^5\otimes \left(\alpha _1 \beta _2\right)+2 i \omega _{16}^5\otimes \left(\alpha _3 \beta
   _1\right)\cr
   &+2 i \omega _{17}^5\otimes \left(\alpha _3 \beta _1\right)-2 i \omega _{20}^5\otimes
   \left(\alpha _1 \beta _1\right)-2 i \omega _{21}^5\otimes \left(\alpha _1 \beta _1\right)-2 i
   \omega _{23}^5\otimes \left(\alpha _2 \beta _3\right)\cr
   &+2 \omega _1^5\otimes \alpha _1+2 \omega
   _2^5\otimes \alpha _1-2 \omega _3^5\otimes \alpha _2+2 \omega _6^5\otimes \alpha _3-2 \omega
   _7^5\otimes \alpha _3\cr
   &+2 \omega _8^5\otimes \alpha _2+5 \omega _{15}^5\otimes \alpha _2+5 \omega
   _{24}^5\otimes \alpha _2+2 \omega _0^5\otimes \beta _3+2 \omega _{11}^5\otimes \beta _2-2 \omega
   _{12}^5\otimes \beta _2\cr
   &-3 \omega _{15}^5\otimes \beta _3+2 \omega _{18}^5\otimes \beta _1-2
   \omega _{19}^5\otimes \beta _1-3 \omega _{24}^5\otimes \beta _3-4 i \omega _5^5\otimes
   \text{I}_4-2 i \omega _{23}^5\otimes \text{I}_4
\end{align}

Notice that displaying these matrices in this tensor product basis allows   displaying about three matrices in the same amount of space where a single $20 \times 20$ matrices could be displayed. This is advantageous when matrices have a certain degree of symmetry, as do the $\brtV_{\rm IJ}$ holoraumy matrices. The $\brL_\rI$ matrices do not have enough symmetry do make displaying them in the tensor product basis particularly advantageous. Nonetheless, representations of all of the $\brL_\rI$, $\brR_\rI$, $\brV_{\rm IJ}$, $\brtV_{\rm IJ}$, $\brV^{(\pm)}_{\rm IJ}$, and the $\brtV^{(\pm)}_{\rm IJ}$ matrices for each of the dWvH, OS, and \nmSG~ multiplets with arbitrary $c_0$, $s_0$, and $g_0$  parameters in both explicit matrix and tensor product form can be found explicitly within or generated from the file \emph{Compare20x20Reps.nb} at the previously mentioned GitHub \href{https://hepthools.github.io/Data/}{data repository}.

\end{document}